\documentclass[aps,prc,twocolumn,showpacs,superscriptaddress,floatfix]{revtex4}

\usepackage{graphicx} 
\usepackage{dcolumn} 
\usepackage{bm} 

\def \FigFactorBig {0.498}
\def \FigFactor {0.4}
\def \FigFactorSm {0.32}
\def \sqrtsNN {\mbox{$\sqrt{s_\mathrm{NN}}$}}
\def \dedx {\mbox{$dE/dx$}}
\def \GeVc {\mbox{$\mathrm{GeV}/c$}}
\def \chisqrndf {\mbox{$\chi^2/ndf$}}
\def \lt {\mbox{$\ <\ $}}
\def \gt {\mbox{$\ >\ $}}
\def \vF {\mbox{$v_4\{\mathrm{EP}_2\}$}}
\def \vO{\mbox{$\lowercase{v_1}\!\left\{\mathrm{EP}_1,\mathrm{EP}_2\right\}$}}
\def \la {\langle}    
\def \ra {\rangle}    
\def \eps {{\varepsilon}}    

\def \ks {$\mathrm{K}^{0}_{S}$ }
\def \llam {$\Lambda + \overline{\Lambda}$ }
\newcommand{\mean}[1]{\left\langle #1 \right\rangle} 
\usepackage{color}

\begin{document}       
       
\title{    
Azimuthal anisotropy in Au+Au collisions at $\sqrtsNN = 200$ GeV }


\affiliation{Argonne National Laboratory, Argonne, Illinois 60439}
\affiliation{University of Bern, 3012 Bern, Switzerland}
\affiliation{University of Birmingham, Birmingham, United Kingdom}
\affiliation{Brookhaven National Laboratory, Upton, New York 11973}
\affiliation{California Institute of Technology, Pasadena, California 91125}
\affiliation{University of California, Berkeley, California 94720}
\affiliation{University of California, Davis, California 95616}
\affiliation{University of California, Los Angeles, California 90095}
\affiliation{Carnegie Mellon University, Pittsburgh, Pennsylvania 15213}
\affiliation{Creighton University, Omaha, Nebraska 68178}
\affiliation{Nuclear Physics Institute AS CR, 250 68 \v{R}e\v{z}/Prague, Czech Republic}
\affiliation{Laboratory for High Energy (JINR), Dubna, Russia}
\affiliation{Particle Physics Laboratory (JINR), Dubna, Russia}
\affiliation{University of Frankfurt, Frankfurt, Germany}
\affiliation{Institute  of Physics, Bhubaneswar 751005, India}
\affiliation{Indian Institute of Technology, Mumbai, India}
\affiliation{Indiana University, Bloomington, Indiana 47408}
\affiliation{Institut de Recherches Subatomiques, Strasbourg, France}
\affiliation{University of Jammu, Jammu 180001, India}
\affiliation{Kent State University, Kent, Ohio 44242}
\affiliation{Lawrence Berkeley National Laboratory, Berkeley, California 94720}
\affiliation{Massachusetts Institute of Technology, Cambridge, Massachusetts 02139}
\affiliation{Max-Planck-Institut f\"ur Physik, Munich, Germany}
\affiliation{Michigan State University, East Lansing, Michigan 48824}
\affiliation{Moscow Engineering Physics Institute, Moscow Russia}
\affiliation{City College of New York, New York City, New York 10031}
\affiliation{NIKHEF, Amsterdam, The Netherlands}
\affiliation{Ohio State University, Columbus, Ohio 43210}
\affiliation{Panjab University, Chandigarh 160014, India}
\affiliation{Pennsylvania State University, University Park, Pennsylvania 16802}
\affiliation{Institute of High Energy Physics, Protvino, Russia}
\affiliation{Purdue University, West Lafayette, Indiana 47907}
\affiliation{University of Rajasthan, Jaipur 302004, India}
\affiliation{Rice University, Houston, Texas 77251}
\affiliation{Universidade de Sao Paulo, Sao Paulo, Brazil}
\affiliation{University of Science \& Technology of China, Anhui 230027, China}
\affiliation{Shanghai Institute of Applied Physics, Shanghai 201800, China}
\affiliation{SUBATECH, Nantes, France}
\affiliation{Texas A\&M University, College Station, Texas 77843}
\affiliation{University of Texas, Austin, Texas 78712}
\affiliation{Tsinghua University, Beijing 100084, China}
\affiliation{Valparaiso University, Valparaiso, Indiana 46383}
\affiliation{Variable Energy Cyclotron Centre, Kolkata 700064, India}
\affiliation{Warsaw University of Technology, Warsaw, Poland}
\affiliation{University of Washington, Seattle, Washington 98195}
\affiliation{Wayne State University, Detroit, Michigan 48201}
\affiliation{Institute of Particle Physics, CCNU (HZNU), Wuhan 430079, China}
\affiliation{Yale University, New Haven, Connecticut 06520}
\affiliation{University of Zagreb, Zagreb, HR-10002, Croatia}

\affiliation{CERN, Switzerland}
\affiliation{INFN, Sez. Di Bari, Bari, Italy}
\affiliation{Instituto de Ciencias Nucleares, UNAM, Mexico}

\author{J.~Adams}\affiliation{University of Birmingham, Birmingham, United Kingdom}
\author{M.M.~Aggarwal}\affiliation{Panjab University, Chandigarh 160014, India}
\author{Z.~Ahammed}\affiliation{Variable Energy Cyclotron Centre, Kolkata 700064, India}
\author{J.~Amonett}\affiliation{Kent State University, Kent, Ohio 44242}
\author{B.D.~Anderson}\affiliation{Kent State University, Kent, Ohio 44242}
\author{D.~Arkhipkin}\affiliation{Particle Physics Laboratory (JINR), Dubna, Russia}
\author{G.S.~Averichev}\affiliation{Laboratory for High Energy (JINR), Dubna, Russia}
\author{S.K.~Badyal}\affiliation{University of Jammu, Jammu 180001, India}
\author{Y.~Bai}\affiliation{NIKHEF, Amsterdam, The Netherlands}
\author{J.~Balewski}\affiliation{Indiana University, Bloomington, Indiana 47408}
% ...

\collaboration{STAR Collaboration}\noaffiliation

\author{A.~Braem}\affiliation{CERN, Switzerland}
\author{M.~Davenport}\affiliation{CERN, Switzerland}
\author{G.~De~Cataldo}\affiliation{INFN, Sez. Di Bari, Bari, Italy}
\author{D.~Di~Bari}\affiliation{INFN, Sez. Di Bari, Bari, Italy}
\author{P.~Martinengo}\affiliation{CERN, Switzerland}
\author{E.~Nappi}\affiliation{INFN, Sez. Di Bari, Bari, Italy}
\author{G.~Paic}\affiliation{Instituto de Ciencias Nucleares, UNAM, Mexico}
\author{E.~Posa}\affiliation{INFN, Sez. Di Bari, Bari, Italy}
\author{F.~Puiz}\affiliation{CERN, Switzerland}
\author{E.~Schyns}\affiliation{CERN, Switzerland}

\collaboration{STAR-RICH Collaboration}\noaffiliation


\date{\today}

\begin{abstract}
The results from the STAR Collaboration on directed flow ($v_1$),
elliptic flow ($v_2$), and the fourth harmonic ($v_4$) in the
anisotropic azimuthal distribution of particles from Au+Au collisions
at $\sqrtsNN = 200$ GeV are summarized and compared with results from
other experiments and theoretical models. Results for identified
particles are presented and fit with a Blast Wave model. Different
anisotropic flow analysis methods are compared and nonflow effects are
extracted from the data. For $v_2$, scaling with the number of
constituent quarks and parton coalescence is discussed. For $v_4$,
scaling with $v_2^2$ and quark coalescence is discussed.
\end{abstract}


\maketitle


\section{\label{sec:intro}Introduction} In heavy-ion collisions at the
Relativistic Heavy Ion Collider (RHIC), the initial
spatially-anisotropic participant zone evolves, via possible novel
phases of nuclear matter, into the observed final state, consisting of
large numbers of produced particles with anisotropic momentum
distributions in the transverse plane.  Important insights into the
evolution may be obtained from the study of this azimuthal anisotropy,
most of which is believed to originate at the early stages of the
collision process.  Unlike at lower beam energies~\cite{review}, the
measured anisotropies at RHIC reach the large values predicted by
hydrodynamic models and conform to the particle mass dependence
expected from hydrodynamics in the kinematic region where this type of
model is expected to be applicable, i.e., for transverse momenta below
a couple of \GeVc~\cite{anonQM04}.  The large observed anisotropy at
RHIC is argued to be indicative of early local thermal equilibrium,
and the particle mass dependence is highly relevant to interpretations
involving a strongly interacting Quark Gluon Plasma
phase~\cite{anonQM04,RikenWSmay04,miklos}. At larger transverse
momenta, measurements of azimuthal anisotropy are also relevant to the
observation of jet
quenching~\cite{STARchargedhighpt,STARhighPtV2Corr}. Given the current
debate around these interpretations, we summarize STAR's findings to
date in the area of azimuthal anisotropy, present additional results
for identified particles, compare in detail the different analysis
methods and their systematic uncertainties, compare the data to
various models, and systematize the results with fits to the
hydrodynamic motivated Blast Wave model.

The paper is organized into sections on the Experiment, Methods of
Analysis, Results, comparison of analysis methods, comparison of
results to various models, and Conclusions. The Methods Comparisons
section is rather technical, dealing with systematic errors, nonflow
effects, and fluctuations.


\section{\label{sec:expt}Experiment} 

\begin{table}[hbt]
\caption{Cuts used in the TPC analysis of Au+Au collisions at
$\sqrtsNN = 200$ GeV. Vertex refers to the event vertex, fit
points are the space points on a track, and dca is the distance of
closest approach of the track to the event vertex.} 
\begin{center}
\begin{tabular}{l l} \hline \hline
  cut     & value              \\ \hline   
  $p_t$   & 0.15 to 2.0 \GeVc  \\   
  $\eta$   & --1.3 to 1.3       \\   
  multiplicity   & \gt 10      \\   
  vertex z   & --25. to 25. cm    \\   
  vertex x, y   & --1.0 to 1.0 cm    \\   
  fit points   & \gt 15         \\   
  fit pts / max. pts   & \gt 0.52  \\   
  dca   & \lt 2.0 cm         \\   
  trigger & min. bias        \\
\hline \hline
\end{tabular}   
\end{center}   
\label{tbl:cuts}   
\end{table}   

The main detectors of the STAR experiment used in these analyses are
the Time Projection Chamber (TPC)~\cite{STARTPC} and the Forward TPCs
(FTPCs)~\cite{STARFTPC}. The Ring Imaging Cherenkov detector
(RICH)~\cite{STARRICH} of the STAR-RICH collaboration is also used for
particle identification. The cuts on the data for most of the TPC
analyses are described in Table~\ref{tbl:cuts}, except for the upper
$p_t$ cutoff which often goes higher as shown in the graphs.  For the
FTPCs the pseudorapidity acceptance is $2.4 \lt |\eta| \lt 4.2$, only
at least 5 hits are required, the distance of closest approach of the
track to the vertex (dca) is restricted to less than 3 cm, and for the
$v_1$ analysis the vertex z is opened up to $\pm 50$ cm.  The RICH
detector~\cite{STARRICH} covers $|\eta| \lt 0.30$ with a $20^{\circ}$
bite in azimuth. The RICH detector separates charged mesons from
protons + anti-protons identified track by track. The admixture of
baryons in the meson sample is always less than 10\%. The momenta of
the particles identified in the RICH come from tracking in the TPC.

\begin{table*}[hbt!]
\caption{Listed for $\sqrtsNN = 200$ GeV for each centrality bin are
the range of the \% most central of the hadronic cross section and its
mean value, the mean charged particle multiplicity with its standard
deviation spread, the estimated mean number of wounded nucleons, the
estimated mean number of binary collisions, and the estimated mean
impact parameter, with the uncertainties in these quantities.}
\begin{center}
\begin{ruledtabular}
\begin{tabular}{l c c c c c c c c c}
Centrality bin & 1 & 2 & 3 & 4 & 5 & 6 & 7 & 8 & 9 \\ \hline
\% most central & 70 -- 80 & 60 -- 70 & 50 -- 60 & 40 -- 50 & 30 -- 40 & 20 --
30 & 10 -- 20 & 5 -- 10 & 0 -- 5 \\
$\mean{centrality} (\%$) & 73.8 & 64.1 & 53.9 & 44.7 & 35.2 & 25.4 & 15.1 &
7.7 & 2.3 \\
$\mean{M}\pm\sigma$ & 38$\pm$11 & 76$\pm$17 & 134$\pm$24 & 214$\pm$32 & 323$\pm$42
 & 468$\pm$53 & 651$\pm$64 & 819$\pm$48 & 961$\pm$56 \\ 
$\mean{N_{WN}}\pm\sigma$ & 13$\pm$4 & 26$\pm$7 & 46$\pm$9 & 75$\pm$11 & 114$\pm$12 & 165$\pm$12 & 232$\pm$10 & 298$\pm$10 & 352$\pm$6 \\
$\mean{N_{binary}}\pm\sigma$ & 11$\pm$5 & 28$\pm$10 & 61$\pm$17 & 120$\pm$28 & 216$\pm$38 & 364$\pm$51 & 587$\pm$61 & 825$\pm$72 & 1049$\pm$72\\
$\mean{b}\pm\sigma$ (fm) & 13.2$\pm$0.6 & 12.3$\pm$0.6 & 11.3$\pm$0.6 & 10.2$\pm$0.5 & 9.0$\pm$0.5 & 7.6$\pm$0.4 & 5.9$\pm$0.3 & 4.2$\pm$0.3 & 2.3$\pm$0.2 \\
\end{tabular}
\end{ruledtabular}
\end{center}
\label{tbl:centrality}
\end{table*}

The data were collected with a minimum bias trigger which required a
coincidence from the two Zero Degree Calorimeters, with each signal
being greater than 1/4 of the single neutron peak and arriving within
a time window centered for the interaction diamond. The centrality
definition, which is based on the raw charged particle TPC
multiplicity with $|\eta| \lt 0.5$, is the same as used
previously~\cite{centrality}. The centrality bins are specified in
Table~\ref{tbl:centrality}.  The mean charged particle multiplicity
given in the Table~\ref{tbl:centrality} is for the cuts in
Table~\ref{tbl:cuts}.  The estimated values in the Table come from a
Monte Carlo Glauber model calculation~\cite{Miller}. In this
calculation, the number of participants is equal to the number of
wounded nucleons. The estimated errors shown for the calculated
quantities come from a linear combination of the changes in the
quantities caused by reasonable variations in the parameters of the
model. Minimum bias refers to 0 to 80\% most central hadronic cross
section.  Two million events are analyzed for this paper. For the
analysis involving FTPCs only 70 thousand events are available.
Errors presented for the data are statistical. Systematic errors are
mainly due to the method of analysis, nonflow effects, and
fluctuations; these will be discussed in Sec.~\ref{sec:methods} on
Methods Comparisons.

Several methods are used to identify particles. The energy loss in the
gas of the TPC identifies particles at low $p_t$. For this the
probability PID method~\cite{STARPID,TangThesis} is used requiring
95\% particle purity unless otherwise stated. In the FTPCs the energy
loss is not sufficient for good particle identification.  The RICH
detector can separate mesons from baryons up to higher $p_t$. Using
the characteristic kink decay of \ks, one is able to go to higher
$p_t$. Strange particles up to high $p_t$ are identified by their
topological decay.

For the kink analysis of charged kaons, flow parameters were measured for
particles that decay in flight within a fiducial volume in the
TPC. The one-prong decay vertex (``kink") provides topological
identification of the particle species with good rejection of
background \cite{kinkMethod}. The main sources of possible
misidentification are pion decays, random combinatoric background, and
secondary hadronic interactions in the TPC gas.  The level of
background in the analyzed sample was estimated to be 5--10\% but is
$p_t$-dependent.  Several cuts were applied to the raw signal in order
to remove most of the background.  Pion decays were removed by
applying a momentum-dependent decay angle cut, which exploits
differences in the decay kinematics.  Other cuts were also applied to
the \dedx, pseudorapidity, and invariant mass of the parent track
candidate, to the daughter momentum, and to the distance of closest
approach associated with the two track segments at the kink vertex.
Finally, there was a quality cut to remove candidates with vertices
inside the TPC sector gaps where spurious kink vertices can arise
\cite{kinkMethod}. Currently the kink method can reconstruct charged
kaons up to $p_t \sim 4$ GeV$/c$.  The tracking software has
difficulty resolving a kink vertex when the decay angle is less than
about $6^\circ$.  For kaons with $p_t \gt 3$ \GeVc, the decay angle is
almost always around $6^\circ$ or less, so efficiency falls off
rapidly above 3 GeV$/c$.  The efficiency also suffers from the limited
fiducial volume; the kaon must decay inside a small sub-volume of the
TPC in order to provide adequate track length for both parent and
daughter tracks.

Other strange particles were identified by their decay
topology~\cite{STARstrange,STARspectra130}. These methods used for the
strange particle decays have already been
described~\cite{STARstrange,STARmultistrange}.


\section{\label{sec:generalMethods}Methods of analysis} Directed and
elliptic flow are defined as the first, $v_1$, and second, $v_2$,
harmonics in the Fourier expansion of the particle azimuthal
anisotropic distribution with respect to the reaction plane. The
reaction plane contains the collision impact parameter. However,
normally measurements are made relative to the observed event plane,
and are corrected for the resolution of the event plane relative to
the reaction plane. The event plane angle is defined for each
harmonic, $n$, by the angle, $\Psi_n$, of the flow vector, $Q$, whose
x and y components are given by
\begin{eqnarray}
Q_n \cos(n \Psi_n) = \sum[w_i \cos(n\phi_i)] \\
Q_n \sin(n \Psi_n) = \sum[w_i \sin(n\phi_i)], \nonumber 
\label{eq:Q}
\end{eqnarray}
where the $\phi_i$ are the azimuthal angles of all the particles used
to define the event plane and the weights, $w_i$, are used to optimize
the event plane resolution. In this paper the weights for the even
harmonics have been taken to be proportional to $p_t$ up to 2 \GeVc\
and constant above that. For the odd harmonics they have been taken to
be proportional to $\eta$ for $|\eta| > 1$.

STAR has previously presented results using different methods of
analysis. In the standard method~\cite{methods}, denoted by $v_n$,
particles are correlated with an event plane of the same
harmonic. Using this method STAR has presented results on elliptic
flow ($v_2$) for charged hadrons~\cite{STARcharged,STARchargedhighpt},
identified particles~\cite{STARPID}, strange
particles~\cite{STARstrange,STAR_Rcp}, and multi-strange
baryons~\cite{STARmultistrange}. In the $N$-particle cumulant
method~\cite{cumulants}, denoted by $v_n\{N\}$, $N$-particle
correlations are calculated and nonflow effects subtracted to first
order when $N$ is greater than 2. Nonflow effects which affect $v_n$
are particle correlations which are not correlated with the reaction
plane. Two-particle cumulants should give essentially the same results
and errors as the standard method, but multi-particle cumulants have
larger statistical errors. STAR has presented four-particle cumulant
results~\cite{STARcumulants} for charged hadrons. In three-particle
mixed harmonic methods relative to the second harmonic event plane,
denoted by $v_n\{\mathrm{EP}_2\}$ when $n \ne 2$, the particles of a
different harmonic are correlated with the well-determined second
harmonic event plane. With mixed harmonics, nonflow effects are
greatly suppressed. With this method STAR has reported results on
directed flow ($v_1$)~\cite{STARv1v4,STARv1,FTPC} and higher harmonics
($v_4$)~\cite{STARv1v4,STARv4} for charged hadrons.


\subsection{\label{sec:v1Methods}Directed flow methods} Because
directed flow goes to zero at midrapidity by symmetry, the first
harmonic event plane is poorly defined in the TPC. A better way to
measure $v_1$ is to use mixed harmonics involving the second harmonic
event plane; this also suppresses nonflow contributions at the same
time. One such method is the three particle cumulant method which has
been described~\cite{v1{3}}.

We also measure $v_1$ using another mixed harmonic technique: we
determine two first order reaction planes $\Psi_1^\mathrm{FTPC_1}$ and
$\Psi_1^\mathrm{FTPC_2}$ in the FTPCs and the second order reaction
plane $\Psi_2^\mathrm{TPC}$ in the TPC. Using the recently proposed
notation (see~\cite{STARv1v4}) we denote this measurement as \vO.
\begin{eqnarray} \label{eq:v1ep1ep2}
v_1\{\mathrm{EP}_1,\mathrm{EP}_2\} = 
\end{eqnarray}
\begin{eqnarray}
\frac{\left\la\cos\left(\phi+\Psi_1^{\mathrm{FTPC}}-2\Psi_2^{\mathrm{TPC}}\right)\right\ra}{\sqrt{\left\la\cos\left(\Psi_1^{\mathrm{FTPC}_1}+\Psi_1^{\mathrm{FTPC}_2}-2\Psi_2^{\mathrm{TPC}}\right)\right\ra\cdot
\mathrm{Res}(\Psi_2^{\mathrm{TPC}})}}\nonumber,
\end{eqnarray}
where the $\phi$ of the particle is correlated with the
$\Psi_1^\mathrm{FTPC}$ in the other subevent, and
\begin{eqnarray}
\mathrm{Res}(\Psi_2^{\mathrm{TPC}}) = \la\cos
[2(\Psi_2-\Psi_{\mathrm{RP}})] \ra
\end{eqnarray}
represents the resolution of the second order event plane measured in
the TPC.  This resolution, as usual, is derived from the square-root
of the correlation of TPC subevent planes. For the derivation of
Eq.~(\ref{eq:v1ep1ep2}) see Appendix~\ref{sec:derivV1}.

This new $v_1$ method also provides an elegant tool to
determine the sign of $v_2$. One of the quantities involved in the
above measurement of \vO\ (see Appendix~\ref{sec:derivV1},
Eq.~(\ref{eq:res}) and compare to \cite{methods}, Eq.~(18)) is
approximately proportional to the product of integrated values of
$v_1^2$ and $v_2$. Applying factors for weights and multiplicities
\cite{methods} leads to
\begin{eqnarray} v_1^2\cdot v_2 \approx
\left(\frac{4}{\pi}\right)^\frac{3}{2} \sqrt{\prod_d
\frac{1}{M_d}\frac{\langle w_d^2\rangle}{\langle w_d\rangle^2}} \\
\times
\left\langle\cos\left(\Psi_1^{\mathrm{FTPC}_1}+\Psi_1^{\mathrm{FTPC}_2}-2\Psi_2^{\mathrm{TPC}}\right)\right\rangle,\nonumber
\end{eqnarray}
where the index $d$ represents the three detectors used
in the analysis: $\mathrm{FTPC}_1$, $\mathrm{FTPC}_2$, and
$\mathrm{TPC}$. For each centrality class $M_d$ denotes the
corresponding multiplicities and $w_d$ are the applied weights
($\eta$-weighting for $\Psi_1$ and $p_T$-weighting for
$\Psi_2$).


\subsection{\label{sec:v2Methods}Elliptic flow methods} The standard
method~\cite{methods} correlates each particle with the event plane
determined from the full event minus the particle of interest. Since
the event plane is only an approximation to the true reaction plane,
one has to correct for this smearing by dividing the observed
correlation by the event plane resolution, which is the correlation of
the event plane with the reaction plane. The event plane resolution is
always less than one, and thus dividing by it raises the flow values.
To make this correction the full event is divided up into two
subevents (a,b), and the square root of the correlation of the
subevent planes is the subevent plane resolution. The full event plane
resolution is then obtained using the equations in Ref.~\cite{methods}
which describe the variation of the resolution with multiplicity.

The scalar product method~\cite{STARcumulants} is a simpler variation
of this method which weights events with the magnitude of the flow
vector $Q$:
\begin{equation} v_n(\eta, p_t) = \frac{ \langle Q_n u_{n,i}^*(\eta,
p_t) \rangle }
  {2 \sqrt{ \langle Q^a_n  {Q^b_n}^* \rangle } } \,,
\label{eq:scalProd}       
\end{equation}
where $u_{n,i}=\cos(n\phi_i) + i \sin(n \phi_i)$ is the unit
vector of the $i^{th}$ particle. If $Q_n$ is replaced by its unit
vector, the above reduces to the standard method. Taking into
account the non-flow contribution, the numerator of
Eq.~(\ref{eq:scalProd}) can be written
as~\cite{STARcumulants,STARhighPtV2Corr}:
\begin{equation}
\la \sum_{i} \cos2(\phi_{p_t}-\phi_i) \ra = M \, v_{2}(p_t) \,
\bar{v}_{2}
+ \{ \text{non-flow}\}
\label{eQAA}
\end{equation}
where $\phi_{p_t}$ is the azimuthal angle of the particle from a given
$p_t$ bin.  The first term in the r.h.s. of Eq.~(\ref{eQAA})
represents the elliptic flow contribution, where $v_2(p_t)$ is the
elliptic flow of particles with a given $p_t$, and $\bar{v}_2$ is the
average flow of particles used in the sum; $M$ is the multiplicity of
particles contributing to the sum, which in this paper is performed
over particles in the region $0.15 \lt p_t \lt 2.0\ \GeVc$ and $|\eta|
\lt 1.0$.

The cumulant method has been well
described~\cite{cumulants,cumuPraticeGuide} and previously used for
the analysis of STAR data~\cite{STARcumulants}.

To reduce the nonflow effects from intra-jet correlations at high
transverse momentum, we also use a modified event plane reconstruction
algorithm, where all subevent particles in a pseudorapidity region of
$|\Delta\eta| \lt$ 0.5 around the highest $p_t$\ particle in the event
are excluded from the event plane determination. With this modified
event plane method, the full event plane resolution is 15--20\% worse
than with the standard method due to the smaller number of tracks used
for the event plane determination.


\subsection{\label{sec:v4Methods}Higher harmonic methods} 
\begin{figure}[!htb]
  \resizebox{\FigFactor\textwidth}{!}{\includegraphics{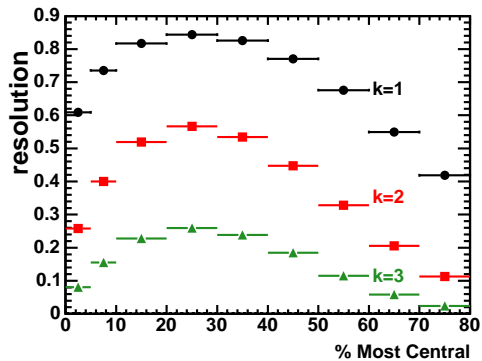}}
  \caption{(color online). The event plane resolutions as a function
  of centrality for $v_{k2}\{\mathrm{EP}_2\}$.
\label{fig:res}}
\end{figure}

Since the
second harmonic event plane is determined so well, one can try to
determine the higher even harmonics of the azimuthal anisotropy by
correlating particles with the second harmonic event plane. However,
then the event plane resolution is worse because of the various
possible orientations of the higher harmonics relative to the second
harmonic event plane. Taking $k$ to be the ratio of the higher
harmonic number to the event plane harmonic number, and using the
equations in Ref.~\cite{methods} we obtain the resolutions in
Fig.~\ref{fig:res} for $v_{k2}\{\mathrm{EP}_2\}$. This method works
when the resolution of the standard method ($k=1$) is large and
therefore those for the higher harmonics are not too low. Also, these
$k \ne 1$ methods use mixed harmonics, which involve multiparticle
correlations, greatly reducing the nonflow contributions.

The cumulant method with mixed harmonics has also been used for
$v_4$~\cite{STARv1v4}.


\section{\label{sec:results}Results} In the following sections we
present results for directed flow, elliptic flow, and the higher
harmonics. Some of the graphs have model calculations on them which
will be discussed in Sec.~\ref{sec:models}. The tables of data for
this paper are available at
\url{http://www.star.bnl.gov/central/publications/}~.


\subsection{\label{sec:v1}Directed flow, \boldmath \vO} 
      
\begin{figure}[!htb]
  \resizebox{\FigFactorBig\textwidth}{!}{\includegraphics{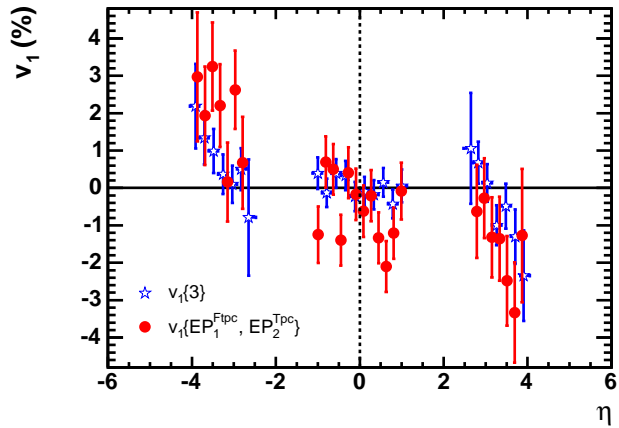}}
  \caption{(color online). Directed flow of charged hadrons as a
  function of pseudorapidity. The measurements of \vO\ (circles;
  centrality 20--60\%) agree with the published results of $v_1\{3\}$
  (stars; centrality 10--70\%).
\label{fig:v1}}
\end{figure}

The STAR TPC has very good capabilities to measure elliptic flow at
mid-rapidity, while the FTPCs allow one to measure directed
flow. Figure~\ref{fig:v1} plots directed flow as a function of
pseudorapidity, showing that $v_1$ appears to be close to zero near
mid-rapidity. First, the analysis was done successfully on simulated
data containing a fixed $v_1$. For real data, using random subevents
in the two FTPCs to determine $\Psi_1^{\mathrm{FTPC}_1}$ and
$\Psi_1^{\mathrm{FTPC}_2}$ in Eq.~(\ref{eq:v1ep1ep2}), the results are
in agreement with the published measurements obtained by the
three-particle cumulant method $v_1\{3\}$~\cite{STARv1v4,STARv1}, as
shown in Fig.~\ref{fig:v1}. Recently, PHOBOS has also
reported~\cite{PHOBOSQM04} $v_1$ values using a two-particle
correlation method. While we approximately agree at $\eta = 4.0$, they
have finite values at $\eta =$ 2.5--3.0, while ours are close to zero,
as can be seen for ours in Fig.~\ref{fig:v1}. 

\begin{figure}[!htb]
  \resizebox{\FigFactor\textwidth}{!}{\includegraphics{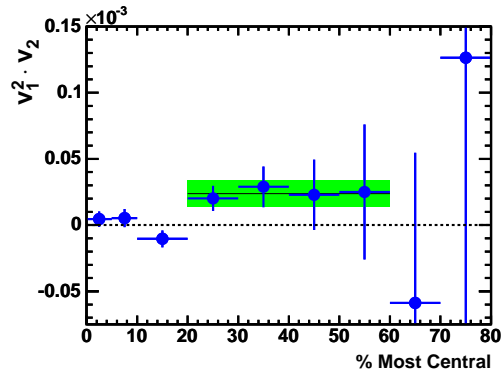}}
  \caption{(color online). The product of $v_1^2$ and $v_2$. The
  shaded band is the mean value of this quantity with its error,
  averaged over centralities 20--60\%. Since this quantity is
  positive, elliptic flow is measured to be {\it in-plane}.
\label{fig:v2Sign}}
\end{figure}

The sign of $v_2$ determines whether the elliptic flow is in-plane or
out-of-plane. Although the sign of $v_2$ had been determined to be
positive from three particle correlations~\cite{STARv1v4}, the above
new method for $v_1$ allows another method based on the sign of
$v_1^2\cdot v_2$. Since $v_1^2$ is always positive, the sign of
$v_1^2\cdot v_2$ determines the sign of $v_2$.

Averaged over centralities 20--60\% we measure $v_1^2 \cdot v_2$ in
Fig.~\ref{fig:v2Sign} to be $(2.38 \pm 0.99) \cdot 10^{-5}$. This is
only a 2.4 sigma effect and if 10\% systematic errors are assumed
based on Sec.~\ref{sec:methods} for both $v_1$ and $v_2$ this becomes
a 2.2 sigma effect. Only the mid-centrality bins are averaged because
in this centrality region the expected nonflow contributions are much
smaller than for the more central and peripheral bins. Therefore, with
these caveats, the sign of $v_2$ is confirmed to be positive: {\it
in-plane} elliptic flow.


\subsection{\label{sec:v2}Elliptic flow, \boldmath $v_2$} There have
been many elliptic flow results from RHIC. STAR has extensive
systematics which we will present and compare to the other
experiments. Many of the graphs will contain Blast Wave model fits
which will be discussed in Sec.~\ref{sec:BW} in Model Comparisons. We
will present data separately for the central rapidity region, the
forward region, and for high $p_t$.
 

\subsubsection{\label{sec:v2Cen}The central region}
\begin{figure}[!htb]
  \resizebox{\FigFactorSm\textwidth}{!}{\includegraphics{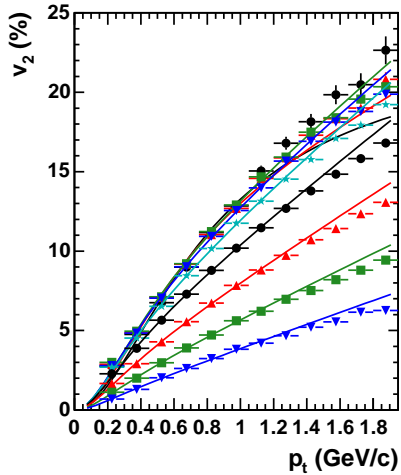}}
  \caption{(color online). Charged hadron $v_2$ vs.\ $p_t$ for the
  centrality bins (bottom to top) 5 to 10\% and in steps of 10\%
  starting at 10, 20, 30, 40, 50, 60, and 70 up to 80\%. The solid
  lines are Blast Wave fits.
\label{fig:v2ptCen}}
\end{figure}

The $v_2(p_t)$ values for charged hadrons for individual centralities
are shown in Fig.~\ref{fig:v2ptCen} with Blast Wave fits performed
assuming that all charged hadrons have the mass of the pion. The data
are well reproduced by the Blast Wave parameterization when $p_t$ is
below 1 \GeVc. Above this limit, the contribution of protons in the
charged hadron sample becomes significant and changes with centrality,
which challenges the pion mass assumption. Furthermore it has been
found that hydrodynamic flow may not be applicable above 1 \GeVc,
especially for light particles, as new phenomena such as hadronization
by recombination may become significant~\cite{reco}.

\begin{figure}[!htb]
  \resizebox{\FigFactor\textwidth}{!}{\includegraphics{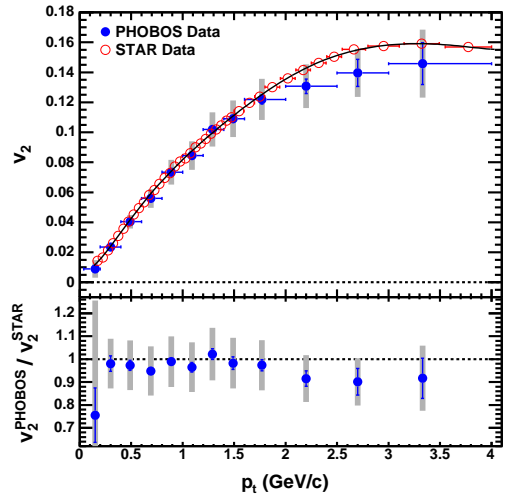}}
  \caption{(color online). $v_2$ vs.\ $p_t$ for charged hadrons from
  0--50\% centrality collisions in comparison to data from
  PHOBOS~\cite{PHOBOSQM04}. The line is a polynomial fit to the STAR
  data. The gray error boxes represent the PHOBOS systematic errors.
  The bottom panel shows the ratio of the PHOBOS data to
  the polynomial fit.
\label{fig:v2Ptphobos}}
\end{figure}

\begin{figure}[!htb]
  \resizebox{\FigFactor\textwidth}{!}{\includegraphics{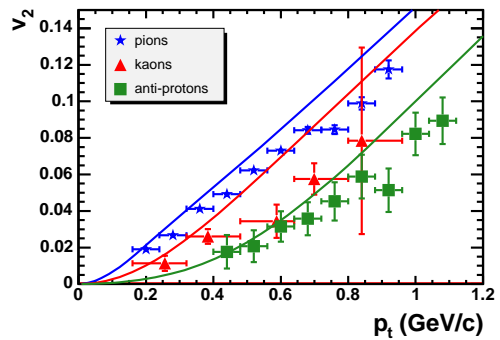}}
  \caption{(color online). $v_2\{4\}$ vs.\ $p_t$ for identified
  particles in the 20--60\% centrality range. The solid lines are
  hydrodynamic calculations~\cite{pasi01}.
\label{fig:v2PID}}
\end{figure}

\begin{figure*}[!htb]
  \resizebox{\FigFactorSm\textwidth}{!}{\includegraphics{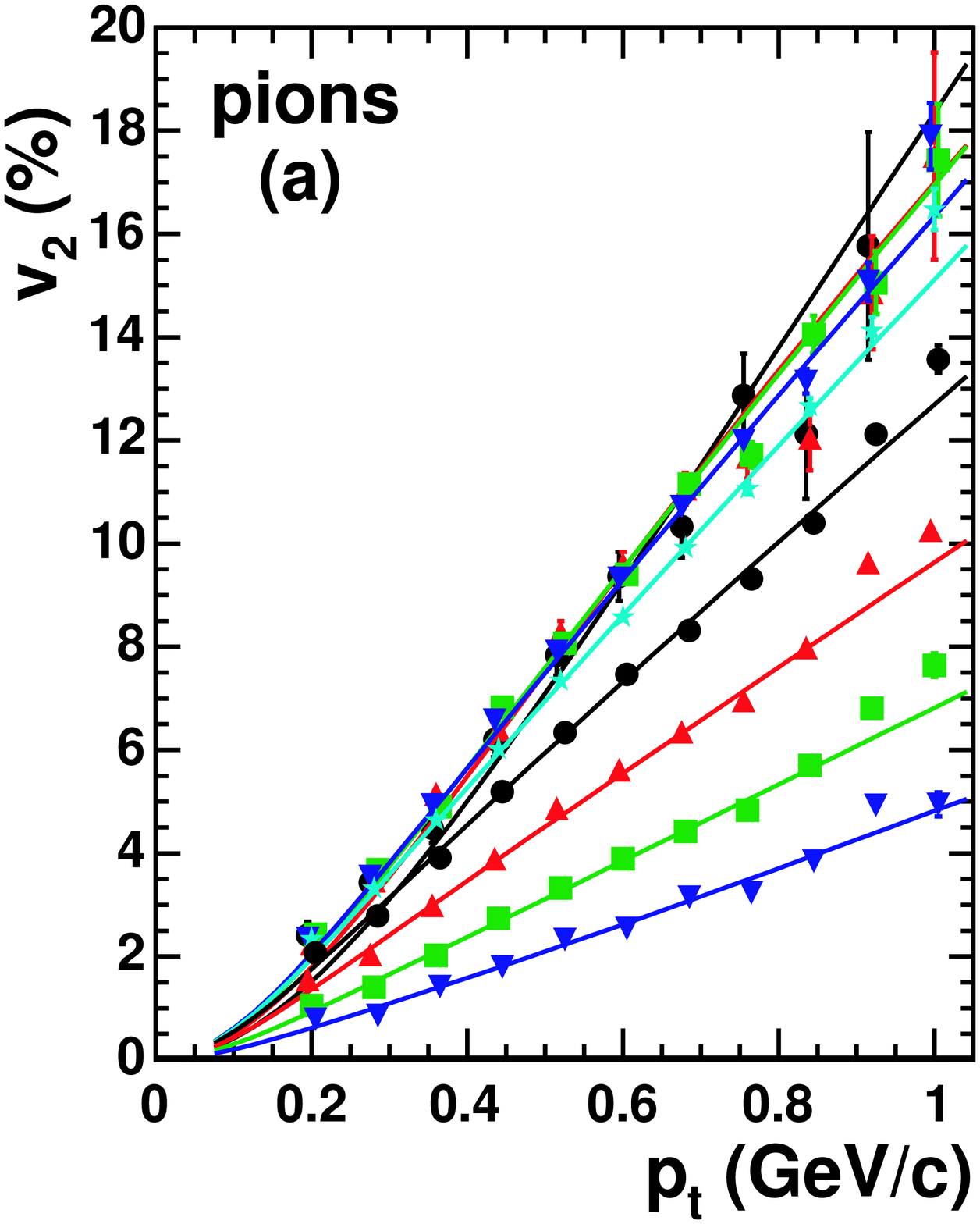}}
  \resizebox{\FigFactorSm\textwidth}{!}{\includegraphics{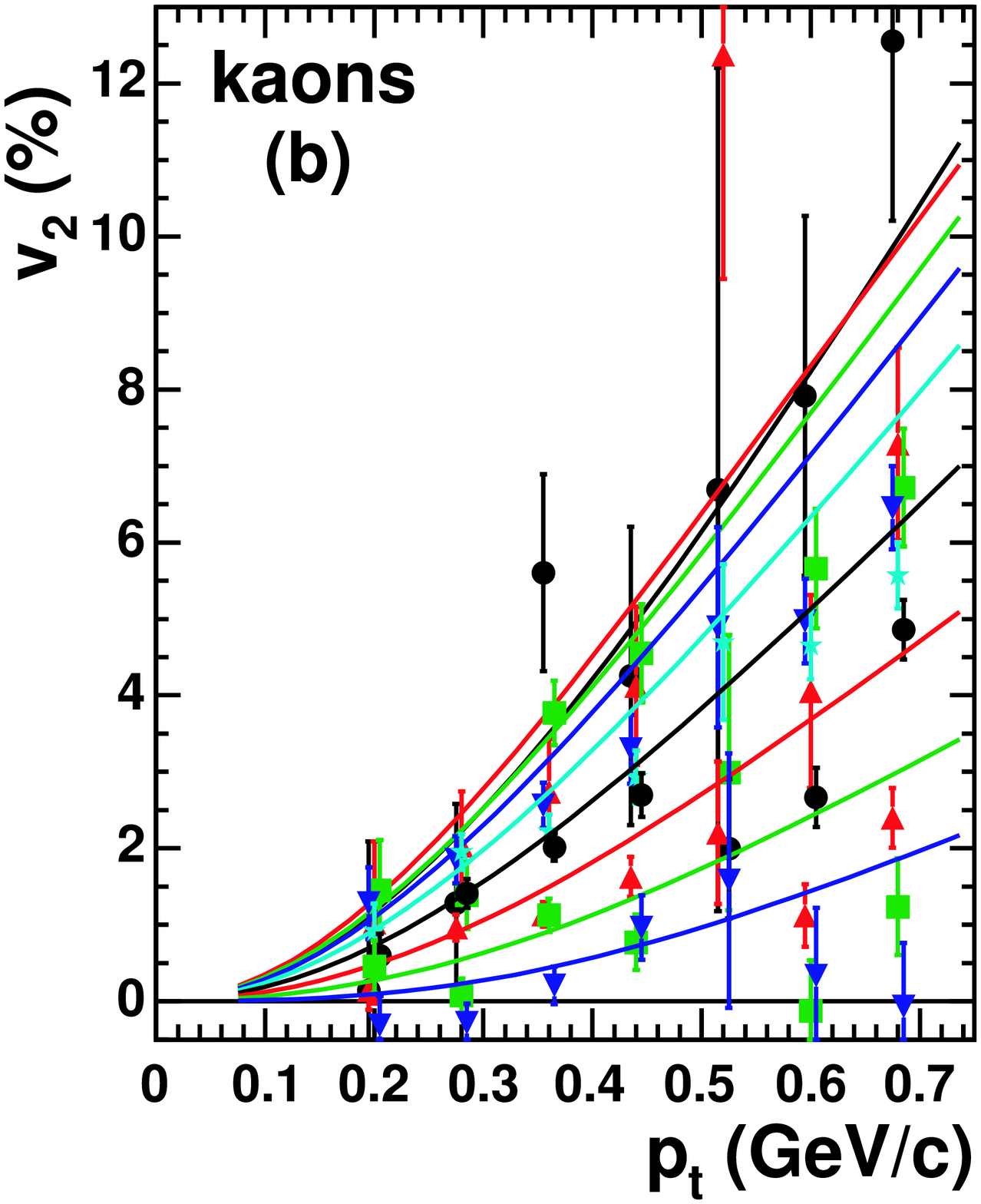}}
  \resizebox{\FigFactorSm\textwidth}{!}{\includegraphics{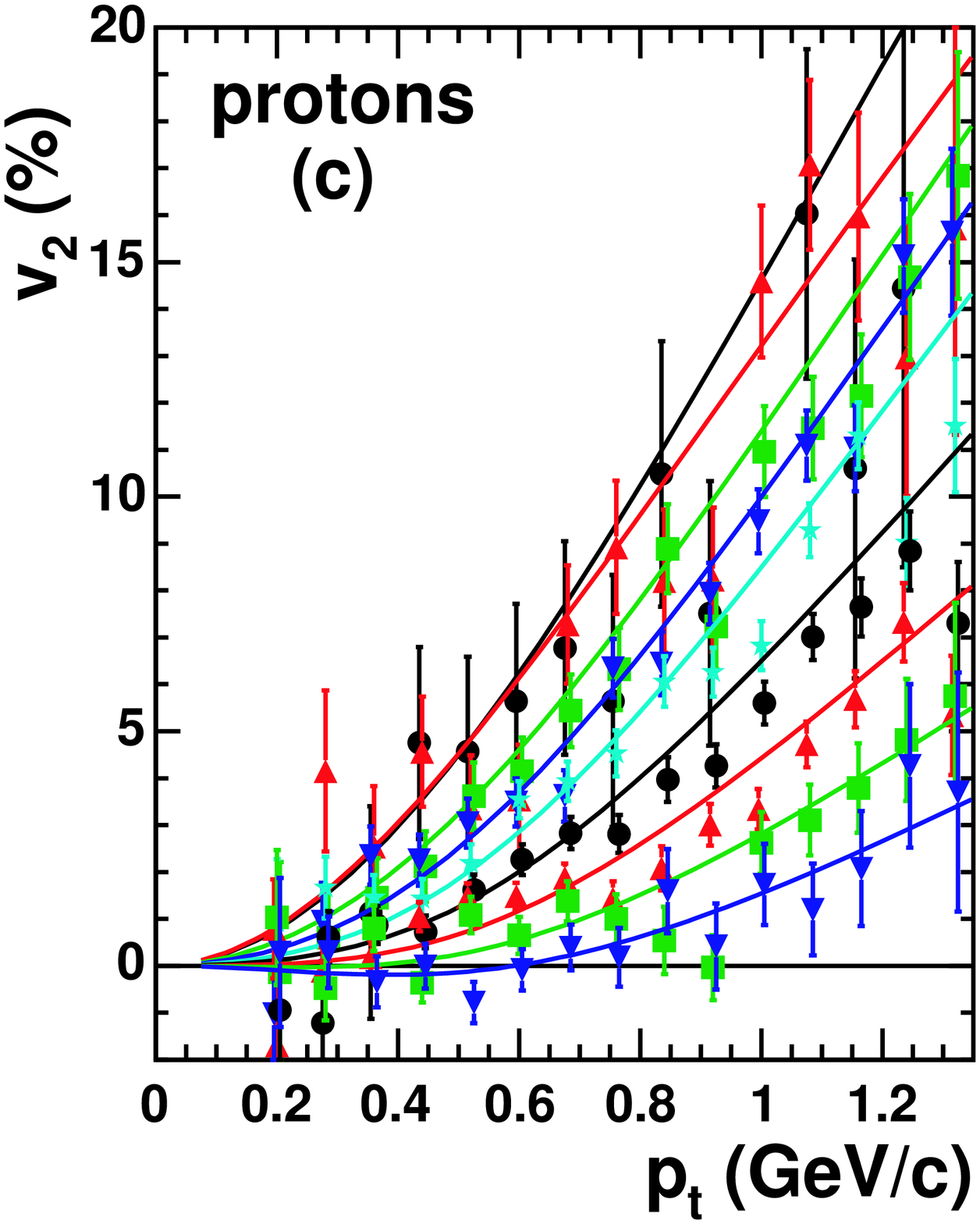}}
  \caption{(color online). $v_2\{2\}$ vs.\ $p_t$ for charged
  pions, charged kaons, and anti-protons for the centrality bins
  (bottom to top) 5 to 10\% and in steps of 10\% starting at 10,
  20, 30, 40, 50, 60, and 70 up to 80\%. The solid lines are Blast
  Wave fits.
\label{fig:v2Pions}}
\end{figure*}

Although all the data presented in this paper were collected using the
full magnetic field (0.5 $T$) of the STAR detector, some data were
also collected using half the magnetic field. Below 0.5 \GeVc\ the
half field $v_2$ values are lower, especially for the more central
collisions. These are regions where the $v_2$ values are small. Adding
the absolute value of 0.0025 to the half field $v_2(p_t)$ data brought
the two sets of data into approximate agreement in this $p_t$
range. This additive value is for both sets of data analyzed with a
dca cut of 2 cm as is done in this paper. The discrepancy gets worse
as the upper dca cut decreases. The effect is not understood and none
of the half field data are included in this paper. However, a possible
explanation is that the half field data have poorer two-track
resolution and are more sensitive to track merging, giving a negative
nonflow contribution. If true, there could be a possible small
residual systematic effect on the full field data.  However, the $v_2$
results are compared to PHOBOS data~\cite{PHOBOSQM04} for 0--50\%
centrality and $0 \lt \eta \lt 1.5$ in Fig.~\ref{fig:v2Ptphobos}. The
STAR data is for the TPC integrated also for 0--50\% centrality. The
full field data presented here agree well with the PHOBOS data.

Results from four-particle cumulants, $v_2\{4\}$, are
shown in Fig.~\ref{fig:v2PID} for particles identified by energy loss
in the TPC. Also shown are hydrodynamic
calculations~\cite{pasi01}. The two-particle values, $v_2\{2\}$, for
pions, kaons, and anti-protons are shown for the individual
centralities with Blast Wave fits in Fig.~\ref{fig:v2Pions}. We
use only anti-protons at low $p_t$ due to contamination of the proton
sample from hadronic interactions in the detector material.

\begin{figure}[!htb]
  \resizebox{\FigFactor\textwidth}{!}{\includegraphics{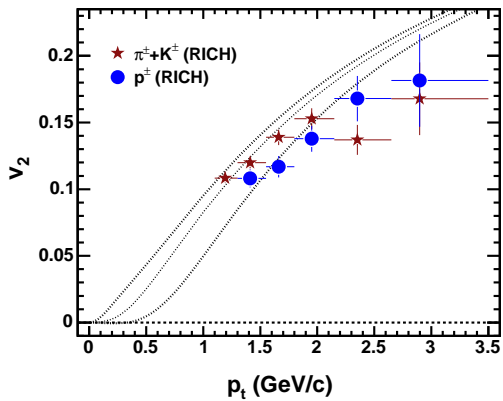}}
  \caption{(color online). $v_2$ vs.\ $p_t$ for particles
  identified in the RICH detector from minimum bias collisions.
  The lines are hydrodynamic calculations~\cite{pasi01} for pions
  (upper line), kaons (middle line), and protons (lower line).
\label{fig:v2RICH}}
\end{figure}

Figure~\ref{fig:v2RICH} shows $v_2(p_t)$ for charged mesons and
protons + anti-protons identified in the RICH detector. The
experimental results are compared to hydrodynamic
calculations~\cite{pasi01}. In the hydrodynamic picture, the mass
ordering of $v_2$ (the lighter particles have larger $v_2$ than the
heavier particles) is predicted to hold at all transverse momenta.  Up
to $p_t \sim 2\ $ \GeVc, $v_2$ of charged mesons is found to be larger
than that of the heavier baryons, in agreement with hydrodynamic
predictions. Above $p_t=2$ \GeVc, the data seem to indicate a reversed
trend where the protons + anti-protons might have larger $v_2$ values
than the charged mesons.

\begin{figure}[!htb]
  \resizebox{\FigFactor\textwidth}{!}{\includegraphics{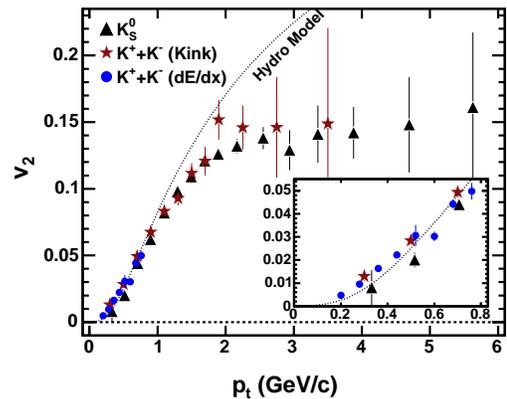}}
  \caption{(color online). $v_2$ vs.\ $p_t$ for neutral and charged
  kaons for minimum bias collisions. The \ks values are from
  Ref.~\cite{STAR_Rcp}. The hydrodynamic model line is from
  Ref.~\cite{pasi01}. The insert expands the low $p_t$ region to
  make the kaons from \dedx\ more visible.
\label{fig:kaon_v2}}
\end{figure}

\begin{figure*}[!htb]
  \resizebox{0.8\textwidth}{!}{\includegraphics{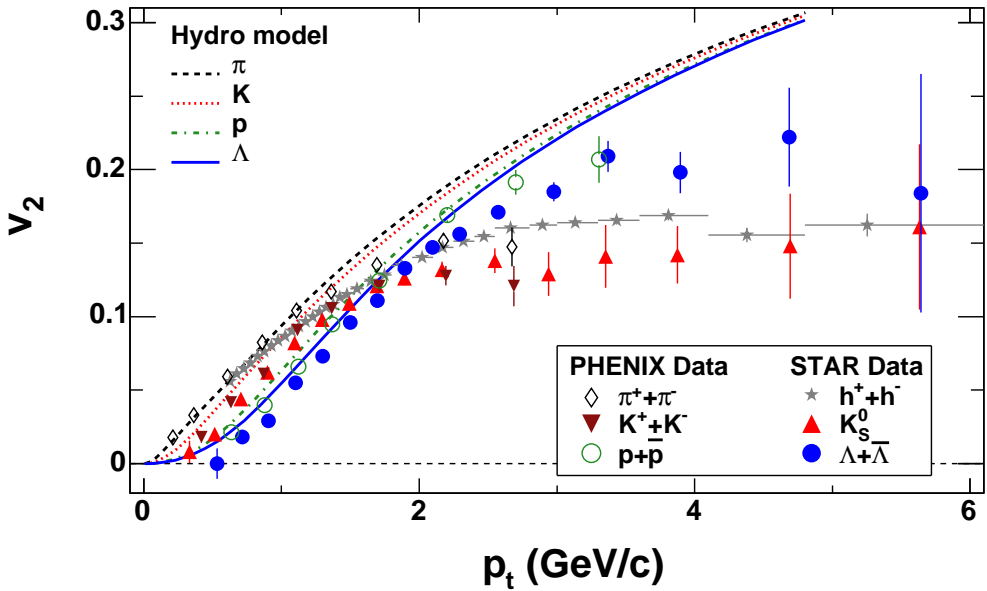}}
  \caption{(color online). $v_2$ vs.\ $p_t$ for strange particles from
  minimum bias collisions. The STAR \ks and \llam values are from
  Ref.~\cite{STAR_Rcp}. The PHENIX data are from Ref.~\cite{PHENIXPID}.
  The hydro calculations are from Ref.~\cite{pasi03}.
\label{fig:v2strange}}
\end{figure*}

From the kink analysis the results are shown in
Fig.~\ref{fig:kaon_v2}. There were about 0.4 accepted candidate kaons
reconstructed per event.

Results are shown in Fig.~\ref{fig:v2strange} comparing STAR data for
\ks\ and \llam\ out to 6 \GeVc\ with some PHENIX
data~\cite{PHENIXPID}, and with hydro calculations~\cite{pasi03}. For
kaons, we can now compare $v_2(p_t)$ for neutral kaons, charged kaons
from kinks, and charged kaons from energy loss identification. This is
shown in Fig.~\ref{fig:kaon_v2}, where the agreement is good, but in
the insert one can see that the neutral kaons tend to be slightly
lower than the charged kaons.

\begin{figure}[!htb]
  \resizebox{\FigFactor\textwidth}{!}{\includegraphics{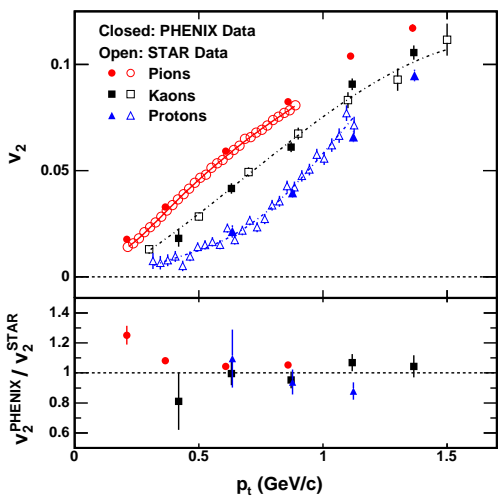}}
  \caption{(color online). $v_2$ vs.\ $p_t$ for charged pions, charged
  kaons, and anti-protons from minimum bias collisions in comparison to
  similar data from PHENIX. The lines are polynomial fits to the STAR
  data. The bottom panel shows the ratio of the PHENIX data to
  the polynomial fits.
\label{fig:phenix}}
\end{figure}

We can also compare our results in more detail at lower $p_t$ with
those from PHENIX~\cite{PHENIXPID}. Figure~\ref{fig:phenix} shows
$v_2(p_t)$ for charged pions and anti-protons from the energy loss
analysis requiring 90\% purity, and kaons from the kink analysis. The
PHENIX results are for $|\eta| \lt 0.35$, for 0--70\% centrality, and
for protons and anti-protons combined. In the $p_t$ range where the
data overlap, the agreement is seen to be good.

It is interesting to see how azimuthal correlations evolve from
elementary collisions (p+p) through collisions involving cold nuclear
matter (d+Au), and then on to hot, heavy-ion collisions (Au+Au). A
convenient quantity for such comparisons is the scalar product.  In
the case of only ``nonflow'', the scalar product should be the same
for all three collision systems regardless of their system size. This
assumes independent collisions and that other effects like short range
correlations are small. Thus, deviations of the scalar product from
elementary p+p collisions result from collective motion and/or effects
of medium modification.

\begin{figure}[!htb]
  \resizebox{\FigFactorBig\textwidth}{!}{\includegraphics{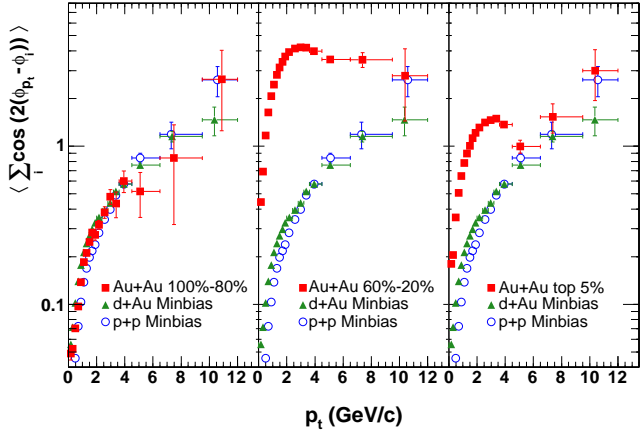}}
  \caption{(color online). Charged hadron azimuthal correlations
  vs.\ $p_t$ in Au+Au collisions (squares) as a function of centrality
  (peripheral to central from left to right) compared to minimum bias
  azimuthal correlations in p+p collisions (circles) and d+Au
  collisions (triangles). The Au+Au and p+p data are from
  Ref.~\cite{STARhighPtV2Corr}.} 
  \label{fig:ppdAuAu} 
\end{figure}
\begin{figure}[!htb]
  \resizebox{\FigFactorBig\textwidth}{!}{\includegraphics{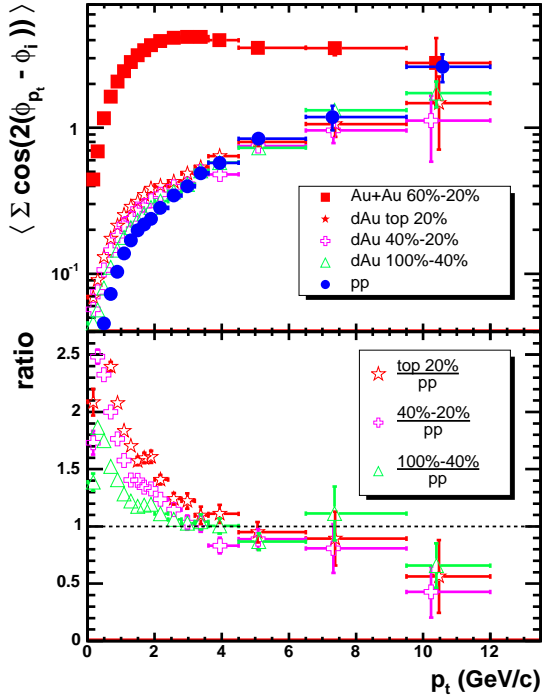}}
  \caption{(color online). The top panel shows the charged
  hadron scalar product vs.\ $p_t$ for different centrality classes
  in d+Au collisions, and minimum bias p+p collisions. The Au+Au
  result is put there for a reference. The bottom panel shows the
  ratio of scalar product from d+Au collisions to minimum bias p+p
  collisions for three different centrality classes.}
\label{fig:ratiodAupp}
\end{figure}

\begin{figure}[!htb]
  \resizebox{\FigFactorSm\textwidth}{!}{\includegraphics{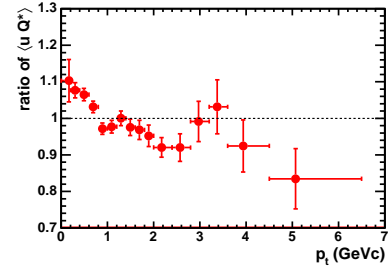}}
  \caption{(color online). The ratio of the scalar product from Au side
  ($-1.0 \lt \eta \lt $$-0.5$) to that from deuteron side ($0.5 \lt \eta
  \lt 1.0$) vs.\ $p_t$ from minimum bias collisions.}
\label{fig:asymdAu}
\end{figure}

Figure~\ref{fig:ppdAuAu} shows the scalar product as defined in
Eq.~\ref{eQAA} as a function of $p_t$ for three different centrality
ranges in Au+Au collisions compared to minimum bias p+p
collisions~\cite{STARhighPtV2Corr} and d+Au collisions. For Au+Au
collisions, in middle central events we observe a big deviation from
p+p collisions that is due to the presence of elliptic flow, while in
peripheral events, collisions are essentially like elementary p+p
collisions. The azimuthal anisotropy goes up to 10 \GeVc\, but we
cannot distinguish whether it is from hydro-like flow or from
jet quenching. For $p_t$ beyond 5 \GeVc\ in central collisions, we
again find a similarity between Au+Au collisions and p+p collisions,
indicating the dominance of nonflow effects. The scalar product in
d+Au collisions is relatively close to that from p+p collisions but
there is a finite difference at low $p_t$. This difference is small if
compared to the difference between middle central Au+Au collisions and
minimum bias p+p collisions. If we examine the difference by looking
in d+Au collisions at different event classes that are defined by the
multiplicity from the Au side (Fig.~\ref{fig:ratiodAupp}), we find
that the scalar product in d+Au increases as a function of
multiplicity class, which is contradictory to Au+Au collisions, in
which the differences rise and fall as a function of centrality; a
typical pattern that is caused by collective flow. The trend in d+Au
could be explained by the Cronin effect, because in high multiplicity
events, the Cronin effect is expected to produce more collective
motion among soft particles in order to generate a high $p_t$
particle~\cite{dAu}. To further test the Cronin effect hypothesis, we
studied the asymmetry of the scalar product in d+Au collisions in
Fig.~\ref{fig:asymdAu}. The ratio of scalar product from the Au side
divided by that from the deuteron side is greater than one at low
$p_t$, and decreases to $\sim$0.9 above $2~\GeVc$. This indicates that
there is more collective motion for $p_t \gt 2~\GeVc$ in the deuteron
side and $p_t \lt 1~\GeVc$ in the Au side, which is again consistent
with the Cronin effect. Recently, the Cronin effect has been explained
by final-state recombination~\cite{Hwa2004}. However the influence of
recombination on azimuthal correlations needs detailed study. In
addition to spectra, the scalar product results open new possibilities
for testing these models.

\subsubsection{\label{sec:v2FTPC}The forward regions}
\begin{figure}[!htb]
  \resizebox{\FigFactor\textwidth}{!}{\includegraphics{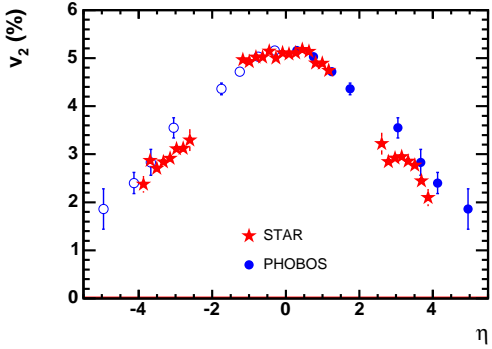}}
  \caption{(color online). Charged hadron $v_{2}$ vs.\ $\eta$ at
  $\sqrtsNN = 200$\,GeV for STAR minimum bias (stars) and
  PHOBOS~\cite{PHOBOSQM04} mid-central (15--25\%) centrality
  (circles). The open circles are PHOBOS data reflected about
  mid-rapidity.}
\label{fig:v2phobos}
\end{figure}

\begin{figure}[!htb]
  \resizebox{\FigFactor\textwidth}{!}{\includegraphics{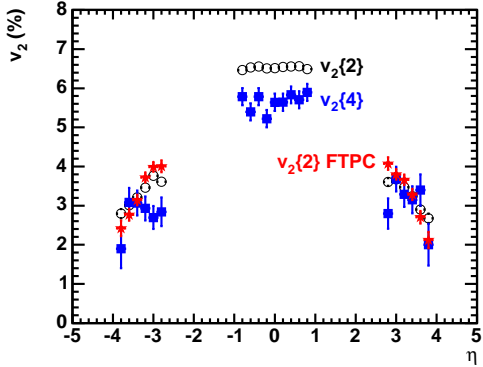}}
  \caption{(color online). Charged hadron $v_2$ vs.\ $\eta$ for
  20--70\% centrality collisions including the FTPC regions. The
  open circles are $v_2\{2\}$, the filled squares are $v_2\{4\}$ and
  the stars are $v_2\{2\}$ for only FTPC particles, not using the main
  TPC particles.}
\label{fig:v2Eta}
\end{figure}

Our measurements of elliptic flow $v_2(\eta)$ for charged hadrons at
forward pseudorapidities along with those from the central region are
shown in Fig.~\ref{fig:v2phobos}. The published
results~\cite{PHOBOS,PHOBOSQM04} obtained by the PHOBOS collaboration
showing a bell-shaped curve are confirmed. We observe a fall-off by a
factor of 1.8 comparing $v_2(\eta = 0)$ with $v_2(\eta =3)$. While
STAR determined the event plane near mid-rapidity, PHOBOS did it at
forward rapidities, which probably accounts for the slightly less
fall-off that they see. Both measurements were done using the standard
method.  Figure~\ref{fig:v2Eta} compares our results for $v_2$
obtained with the method of two-particle cumulants, $v_2\{2\}$, to
that for four-particle cumulants, $v_2\{4\}$. The difference at
mid-rapidity will be discussed in Sec.~\ref{sec:methods}. The FTPC
$v_2\{4\}$ values are not quite symmetric about mid-rapidity, but not
unreasonable considering the statistical errors. Within the errors in
the FTPC regions, the values from the different methods are about the
same.

\begin{figure}[!htb]
  \resizebox{\FigFactor\textwidth}{!}{\includegraphics{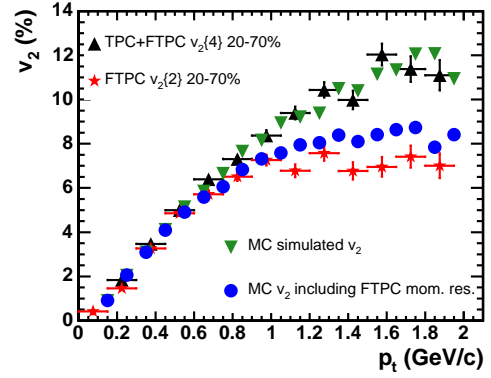}}
  \caption{(color online). Charged hadron $v_2\{2\}$ vs.\ $p_t$
  for centrality 20--70\% in the FTPC ($2.5 \lt |\eta| \lt 4.0$)
  regions (stars) compared to $v_2\{4\}$ in the TPC + FTPCs
  (up-triangles). The down-triangles are Monte Carlo results fit to
  the up-triangles, and the solid circles include the FTPC momentum 
  resolution.}
\label{fig:v2FTPCpt}
\end{figure}

Figure~\ref{fig:v2FTPCpt} shows $v_2\{4\}(p_t)$ obtained from the
four-particle cumulant method. Since there are many more particles in
the main TPC than in the FTPCs, these $v_2$ values are mainly at
mid-rapidity. $v_2\{4\}$, which is much less sensitive to nonflow
effects, is compared to $v_2\{2\}$ at forward rapidities, where
nonflow may be small. The observed flattening at $p_t$ values around 1
\GeVc\ for the FTPC measurements might be explained by the momentum
resolution of the FTPCs. To quantify the influence of the momentum
resolution a Monte-Carlo simulation of $v_2(p_t)$ based on the
measurements at mid-rapidity was done, but the input $\eta$ and $p_t$
spectra were obtained from measurements of the Au+Au minimum bias data
at forward rapidities. Results of embedding charged pions (neglecting
protons) in real Au+Au events up to 5\% of the total multiplicity in
the FTPCs were used to estimate the momentum resolution as a function
of $\eta$ and $p_t$. At $\eta = 3.0$ the momentum resolution goes from
10\% at low $p_t$ to 35\% at $p_t = 2.0\ \GeVc$, but gets about a
factor of two worse at $\eta = 3.5$. In Fig.~\ref{fig:v2FTPCpt} the MC
simulation $v_2(p_t)$ including the momentum resolution of the FTPCs
seems to explain the observed flattening by smearing low $p_t$
particles to higher $p_t$. Thus we can not conclude that the shape of
the $p_t$ dependence of elliptic flow at forward rapidities is
different from that at mid-rapidity, even though the values integrated
over $p_t$ are considerably smaller as shown in Fig.~\ref{fig:v2Eta}.


\subsubsection{High $p_t$} Hadron yields at sufficiently high
transverse momentum in Au+Au collisions are believed to contain a
significant fraction originating from the fragmentation of high energy
partons resulting from initial hard scatterings. Calculations based on
perturbative QCD predict that high energy partons traversing nuclear
matter lose energy through induced gluon radiation~\cite{quenching}.
Energy loss (jet quenching) is expected to depend strongly on the
color charge density of the created system and the traversed path
length of the propagating parton. Consistent with jet quenching
calculations, strong suppression of the inclusive high-$p_t$ hadron
production~\cite{centrality,highptsuppression} and back-to-back
high-$p_t$ jet-like correlation~\cite{btob} compared to the reference
p+p and d+Au systems was measured in central Au+Au collisions at RHIC.
In non-central heavy-ion collisions, the geometrical overlap region
has an almond shape in the transverse plane, with its short axis lying
in the reaction plane. Partons traversing such a system, on average,
experience different path-lengths and therefore different energy loss
as a function of their azimuthal angle with respect to the reaction
plane. This leads to an azimuthal anisotropy in particle production at
high transverse momenta. Finite values of $v_2$ were measured in
non-central Au+Au collisions for $p_t$ up to $\sim$ 7--8 \GeVc\
\cite{STARchargedhighpt,STARhighPtV2Corr} using the standard reaction
plane method and two- and four-particle cumulants. The measurements of
azimuthal anisotropies at high transverse momenta with the standard
reaction plane method and two-particle cumulants are influenced by the
contribution from the inter- and intra-jet correlations. These
correlations, in general, may not be related to the true reaction
plane orientation and, hence, are a source of nonflow effects. A
multi-particle cumulant analysis, which has been shown to suppress
nonflow effects, may give lower $v_2$ values because of the opposite
sensitivity of $v_2\{2\}$ and $v_2\{4\}$ to the fluctuations of $v_2$
itself described in Sec.~\ref{sec:nonflow} and Ref.~\cite{epsilonMC}.

\begin{figure}[!htb]
\resizebox{\FigFactor\textwidth}{!}{\includegraphics{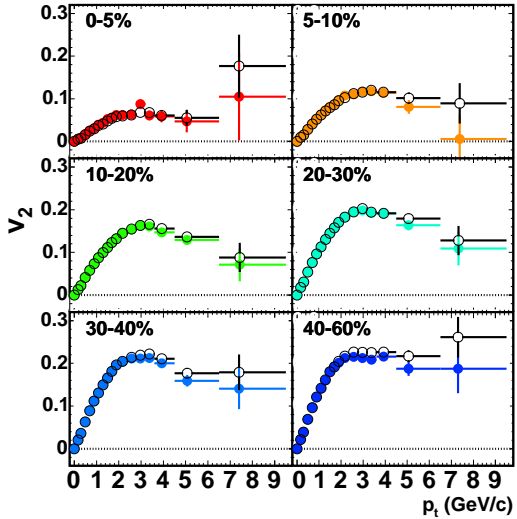}}
\caption{(color online). $v_2$\ vs.\ $p_t$\ for charged hadrons for 
different centrality bins. The standard reaction plane method is shown
by open symbols and the modified reaction plane method by solid
symbols.
\label{fig:v2method}}
\end{figure}

\begin{figure}[!htb]
\resizebox{\FigFactor\textwidth}{!}{\includegraphics{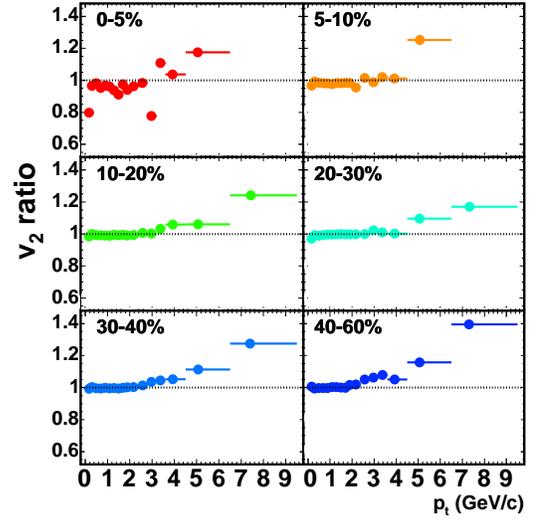}}
\caption{(color online). $v_2$\ for the standard reaction plane method
divided by $v_2$ for the modified reaction plane method vs.\ $p_t$ for
charged hadrons in different centrality bins. Error bars are not shown
as the same dataset is used for both methods.
\label{fig:v2methodratio}} 
\end{figure} 

\begin{figure}[!htb]
\resizebox{\FigFactor\textwidth}{!}{\includegraphics{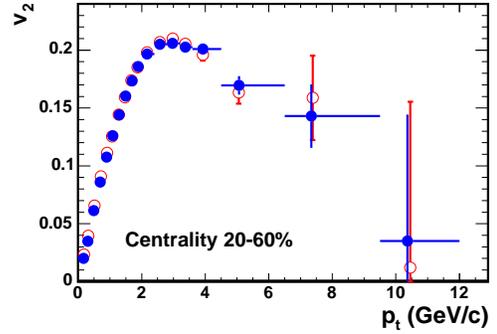}}
\caption{(color online). $v_2$ vs.\ $p_t$\ for charged hadrons from
the modified reaction plane method (solid circles). Open circles (from
Ref.~\cite{STARhighPtV2Corr}, Fig.~2) are the two-particle cumulant
results after subtracting the correlations measured in p+p collisions.
Error bars show statistical uncertainties only.
\label{fig:v2_20-60}}
\end{figure}

Figure~\ref{fig:v2method} shows the differential elliptic flow $v_2$
obtained with the standard and modified reaction plane methods as a
function of $p_t$\ for different collision centralities. The modified
event plane method excludes particles within $|\Delta\eta| \lt$ 0.5
around the highest $p_t$\ particle. For both methods $v_2$ rises
linearly up to $p_t = 1$ \GeVc, then deviates from a linear rise and
saturates for $p_t \gt $3 \GeVc\ for all
centralities. Fig.~\ref{fig:kaon_v2} shows a similar behavior.
Although the statistical errors are large, we observe a systematic
difference in Fig.~\ref{fig:v2method} for the $v_2$ values obtained
with the two methods at high transverse momenta.  This is better
illustrated in Fig.~\ref{fig:v2methodratio}, where we show the ratio
of $v_2$ obtained with the standard and modified reaction plane
methods. At low transverse momenta ($p_t \lt$ 2 \GeVc), the $v_2$
values are very similar for both methods. At higher transverse
momenta, $v_2$ is systematically larger for the standard reaction
plane method. For more peripheral collisions this effect is larger,
and it also begins at lower $p_t$. The modified reaction plane method
seems to eliminate at least some of the nonflow effects at high
transverse momenta (up to 15--20\% at $p_t =$ 5--6 \GeVc\ in the most
peripheral collisions). The contribution of the azimuthal correlations
not related to the reaction plane orientation has been previously
studied using p+p collisions~\cite{STARhighPtV2Corr}. In p+p
collisions, all correlations are considered to be of nonflow
origin. In Fig.~\ref{fig:ppdAuAu} the azimuthal correlations in
mid-central Au+Au collisions are very different from those in p+p
collisions in both magnitude and $p_t$
dependence. Figure~\ref{fig:v2_20-60} shows the modified reaction
plane results on $v_2(p_t)$ for charged hadrons of centrality
20--60\%.  We find a very good agreement of $v_2$ from the modified
reaction plane analysis with the two-particle cumulant results after
subtracting the correlations measured in p+p
collisions~\cite{STARhighPtV2Corr}. Neither of these modified methods
which seem to be necessary at high $p_t$\ give results which differ
from the simple standard method below $p_t$\ of 2 \GeVc, and thus are
not used in the other analyses of this paper.


\subsection{\label{sec:v4}Higher harmonics}


\subsubsection{\label{sec:v4Cen}The central region}
\begin{figure}[!htb]
  \resizebox{\FigFactor\textwidth}{!}{\includegraphics{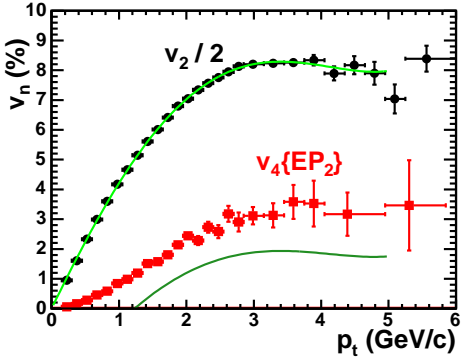}}
  \caption{(color online). $v_2$ scaled down by a factor of 2, and
  \vF\ vs.\ $p_t$ for charged hadrons from minimum bias
  events. Using a fit to the $v_2$ values, the lower solid line is the
  predicted $v_4$ needed to just remove the ``peanut'' waist (see
  text).
\label{fig:waist}}
\end{figure}

\begin{figure}[!htb]
  \resizebox{\FigFactorSm\textwidth}{!}{\includegraphics{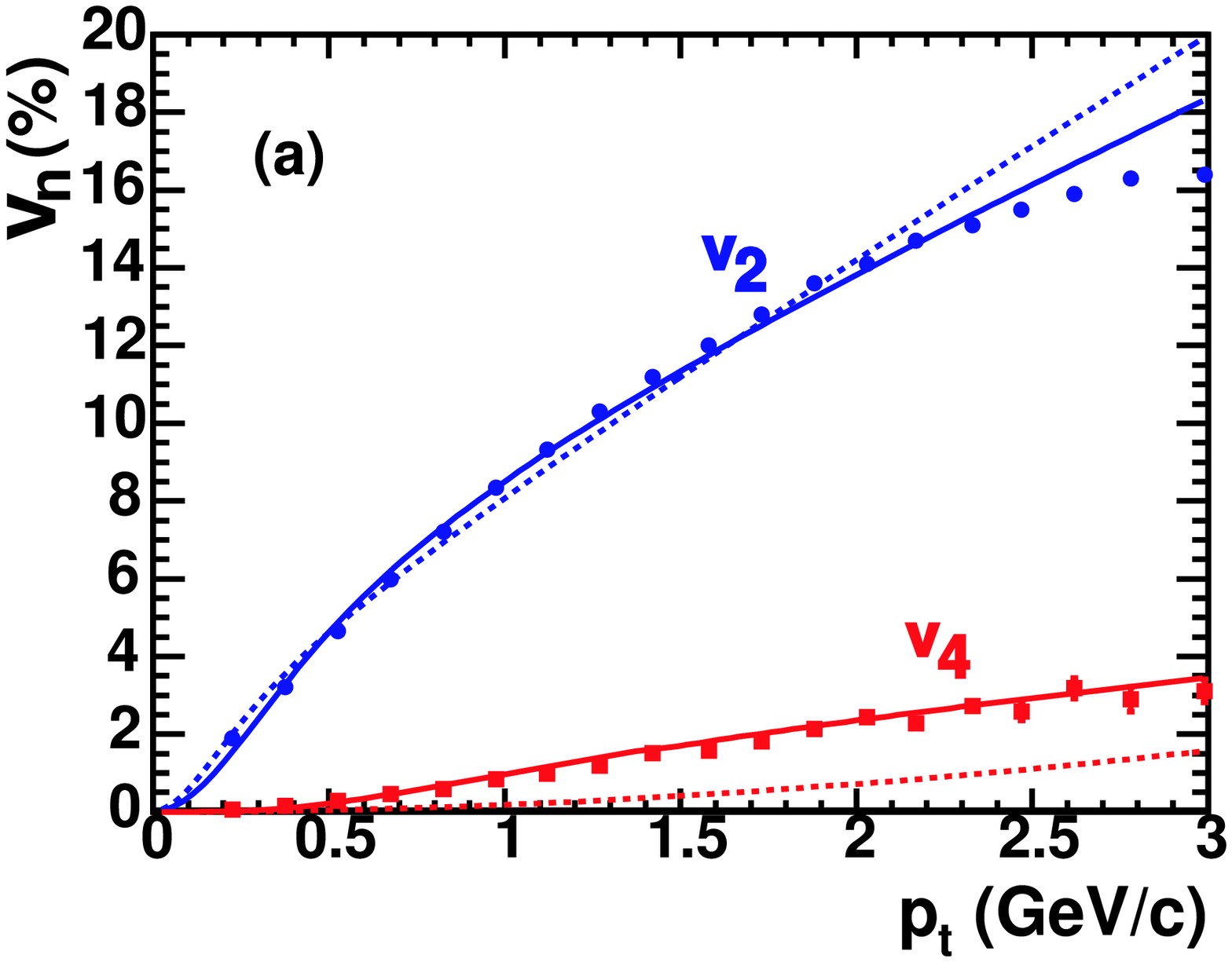}}
  \resizebox{\FigFactorSm\textwidth}{!}{\includegraphics{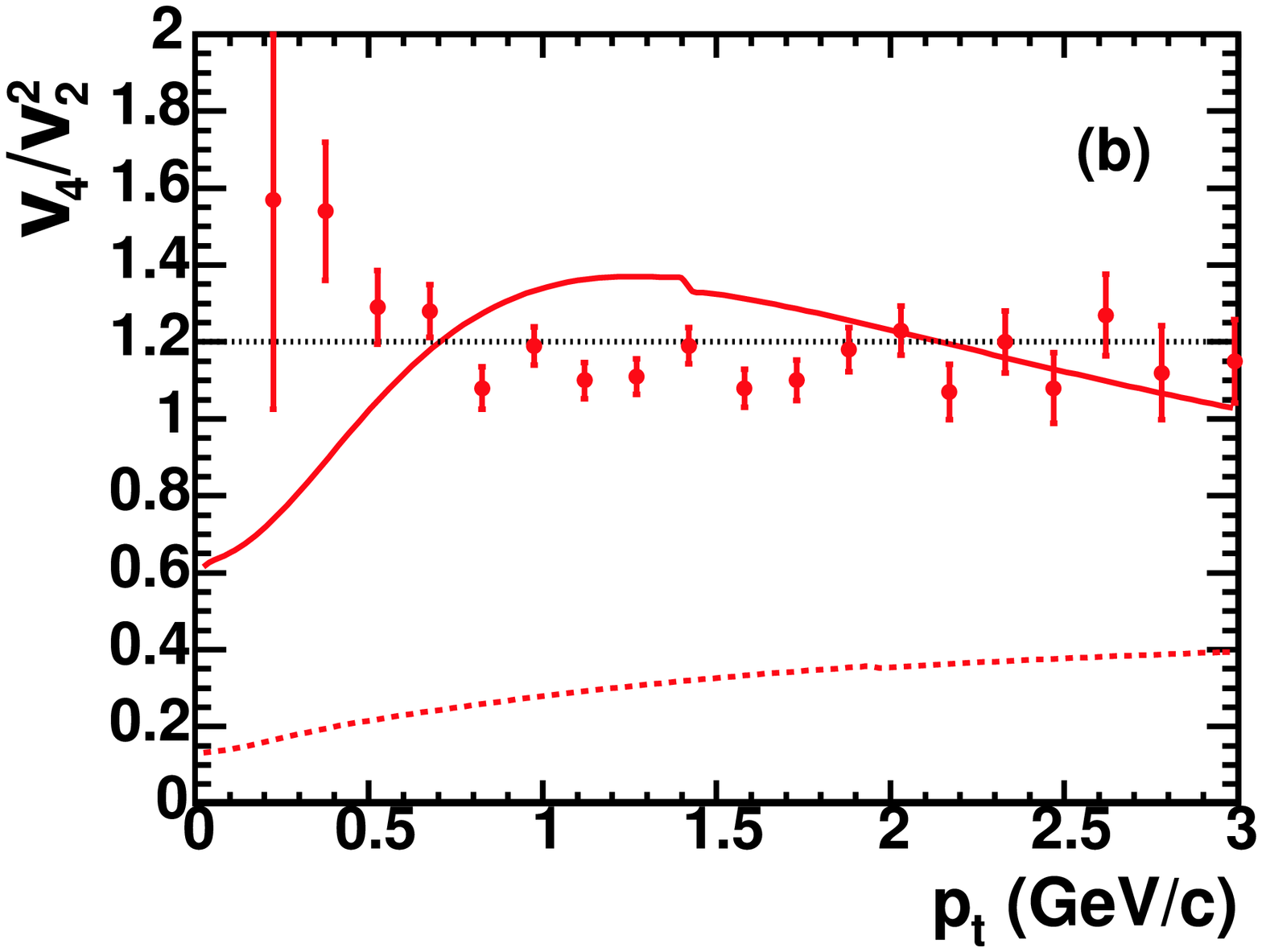}}
  \caption{(color online). Graphs of $v_n$ and $v_4/v_2^2$. The dashed
  lines are surface shell Blast Wave fits with no $\rho_4$ or
  $s_4$ terms (See Sec.~\ref{sec:BW}) to the charged hadron $v_2$
  minimum bias data. The resultant ratio $v_4/v_2^2$ is shown as the
  lower dashed line in the ratio graph (b). The solid lines are the fits
  with the addition of $\rho_4$ and $s_4$. The resultant ratio
  $v_4/v_2^2$ is shown as the solid curve in the ratio graph (b). The
  dotted line in the ratio graph (b) at 1.2 represents the average value
  of the data.
\label{fig:BW}}
\end{figure}

Our results for charged hadron $v_4$ and $v_6$ from this study have
already been published~\cite{STARv1v4,STARv4}, and $v_4(p_t)$ is shown
again in Fig.~\ref{fig:waist}. It also was found that $v_4$ scales as
$v_2^2$. The value of $v_4/v_2^2$ was found to be 1.2, almost
independent of $p_t$~\cite{STARv4}, as can be seen in the ratio graph
of Fig.~\ref{fig:BW}~(b).

\begin{figure}[!htb]
  \resizebox{\FigFactorSm\textwidth}{!}{\includegraphics{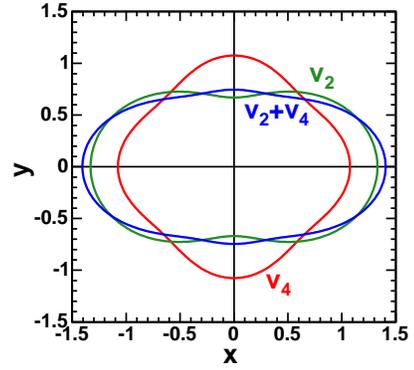}}
  \caption{(color online). A polar graph of the distribution $1 +
  2 v_2 \cos(2\phi) + 2 v_4 \cos(4\phi)$ where $\phi$ is the
  azimuthal angle relative to the positive $x$ axis. Plotted are the
  distributions for $v_2 = 16.5$\% showing the waist, $v_4 = 3.8$\%
  having a diamond shape, and both coefficients together.
\label{fig:polar}}
\end{figure}

\begin{figure}[!htb]
  \resizebox{\FigFactorSm\textwidth}{!}{\includegraphics{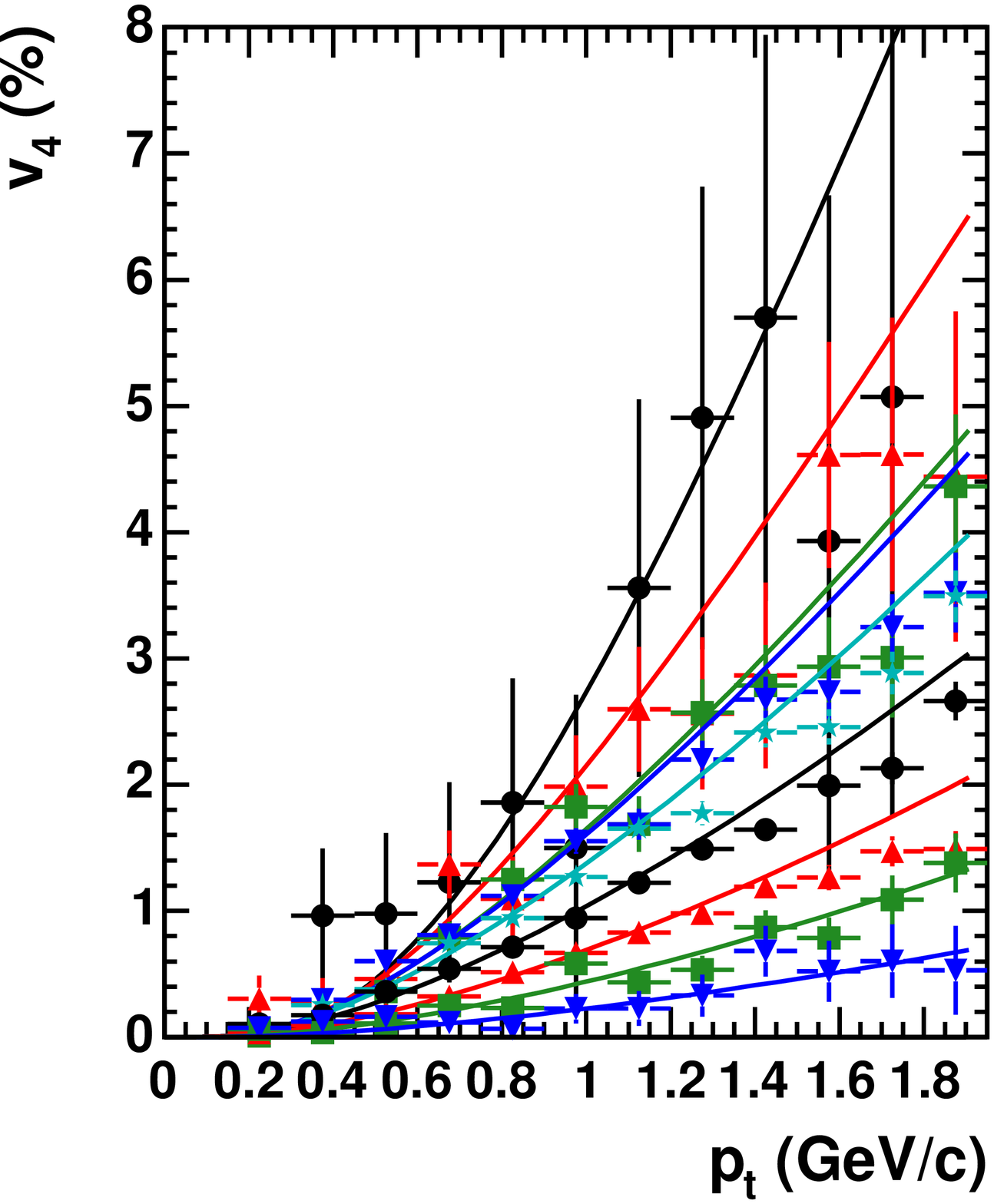}}
  \caption{(color online). $\vF$ vs.\ $p_t$ for charged hadrons for
  the centrality bins (bottom to top) 5 to 10\% and in steps of 10\%
  starting at 10, 20, 30, 40, 50, 60, and 70 up to 80\%. The solid
  lines are Blast Wave fits.  
\label{fig:v4ptCen}} 
\end{figure}

\begin{figure*}[!htb]
  \resizebox{\FigFactor\textwidth}{!}{\includegraphics{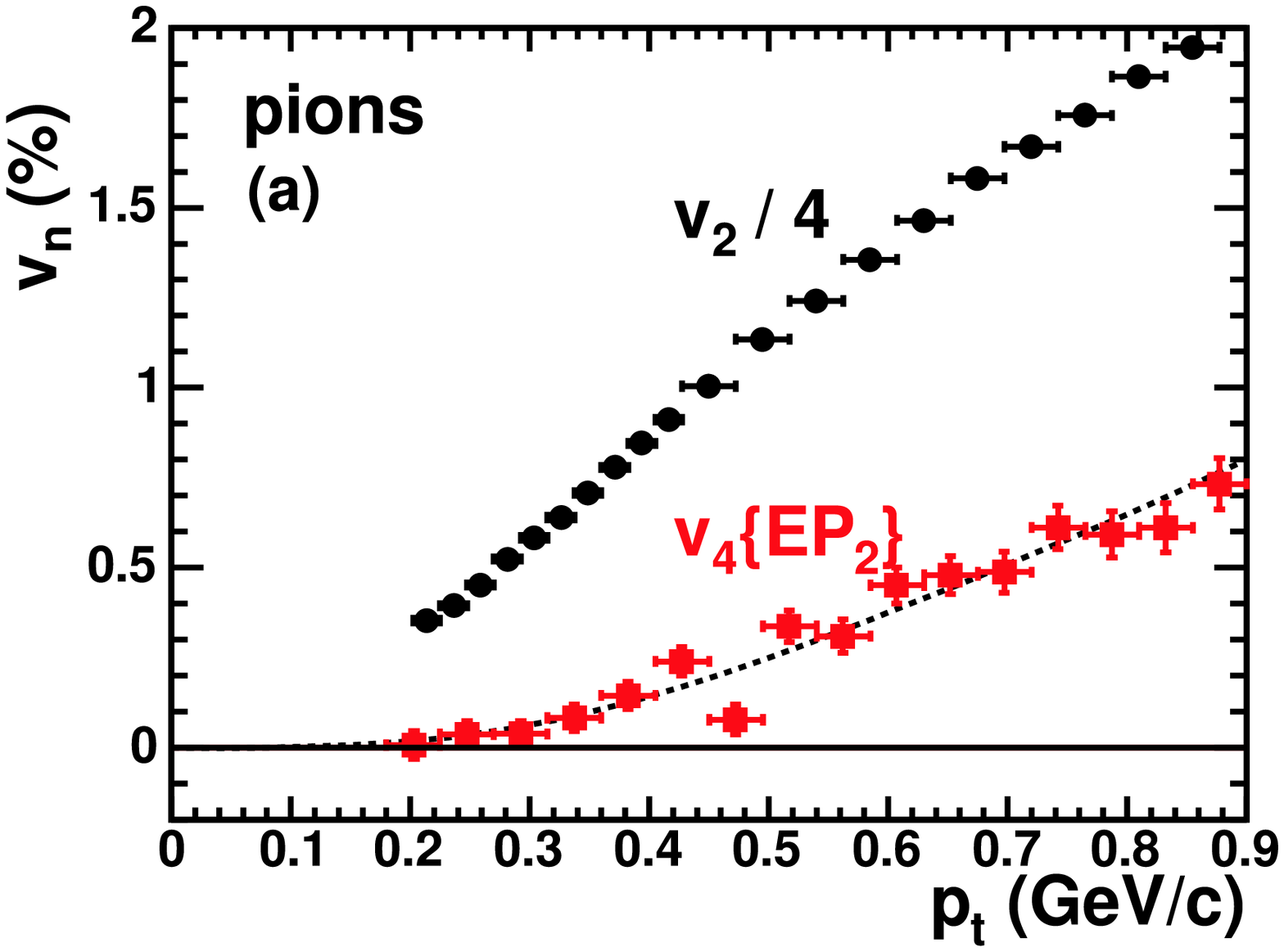}}
  \resizebox{\FigFactor\textwidth}{!}{\includegraphics{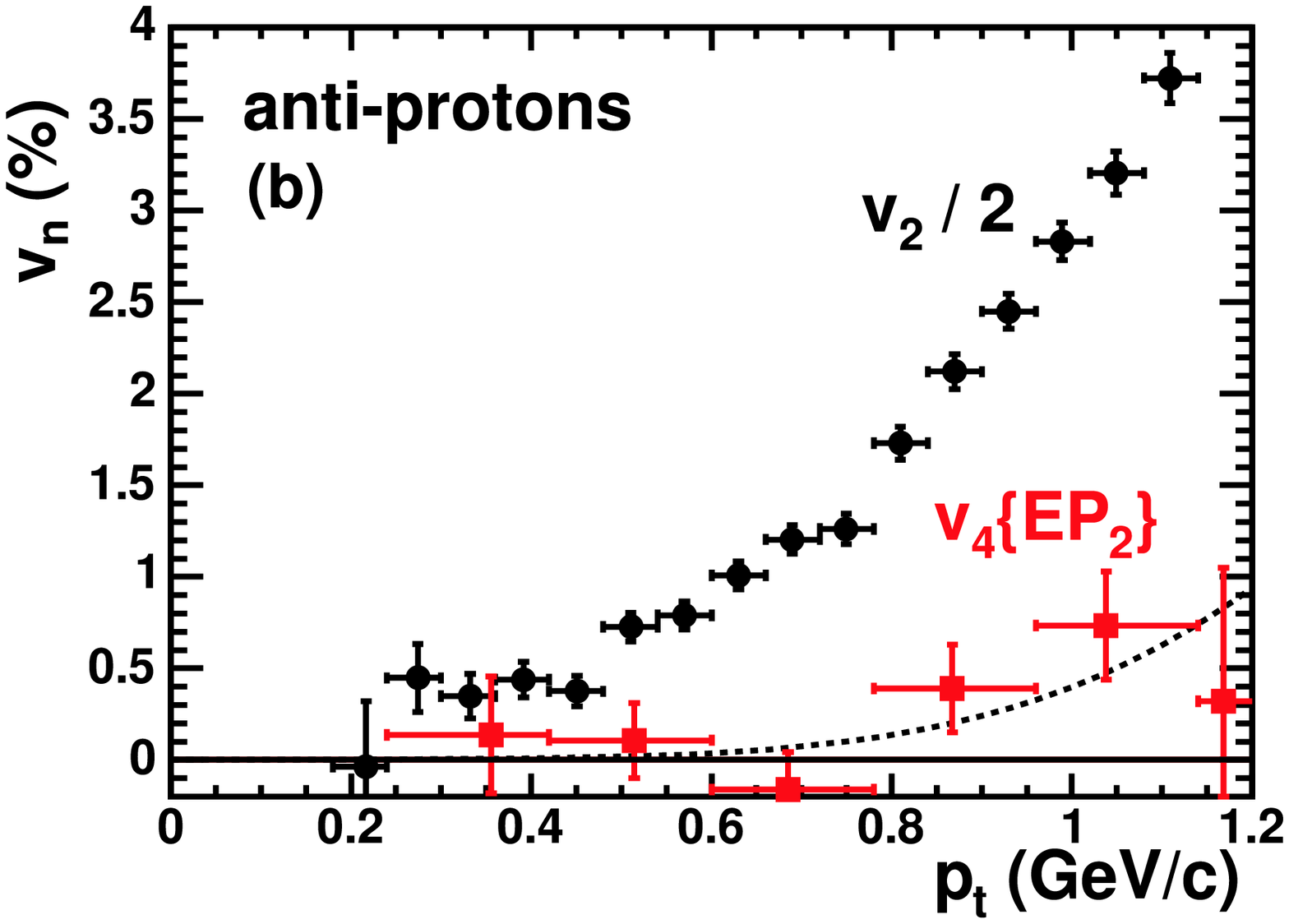}}
  \resizebox{\FigFactor\textwidth}{!}{\includegraphics{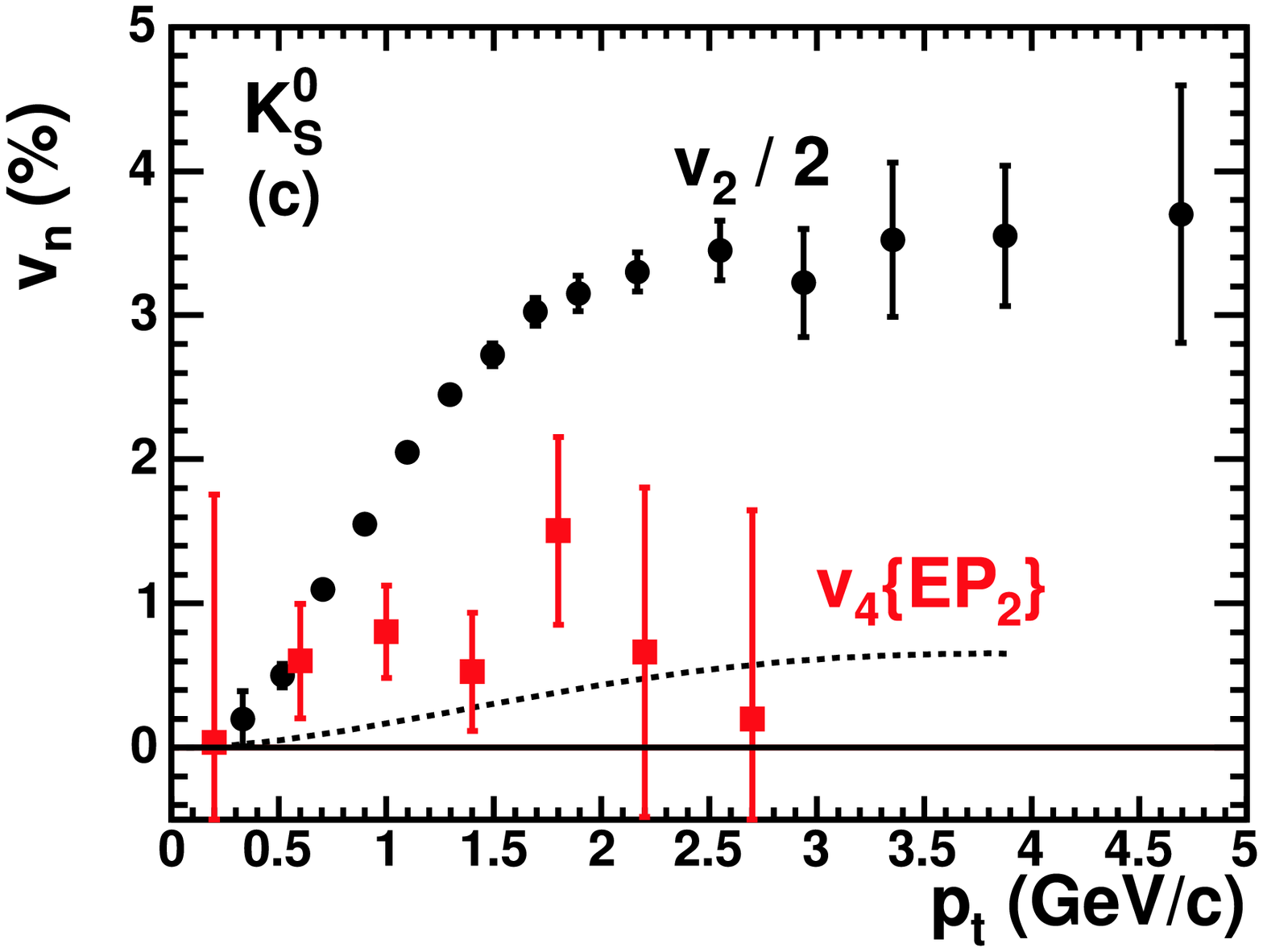}}
  \resizebox{\FigFactor\textwidth}{!}{\includegraphics{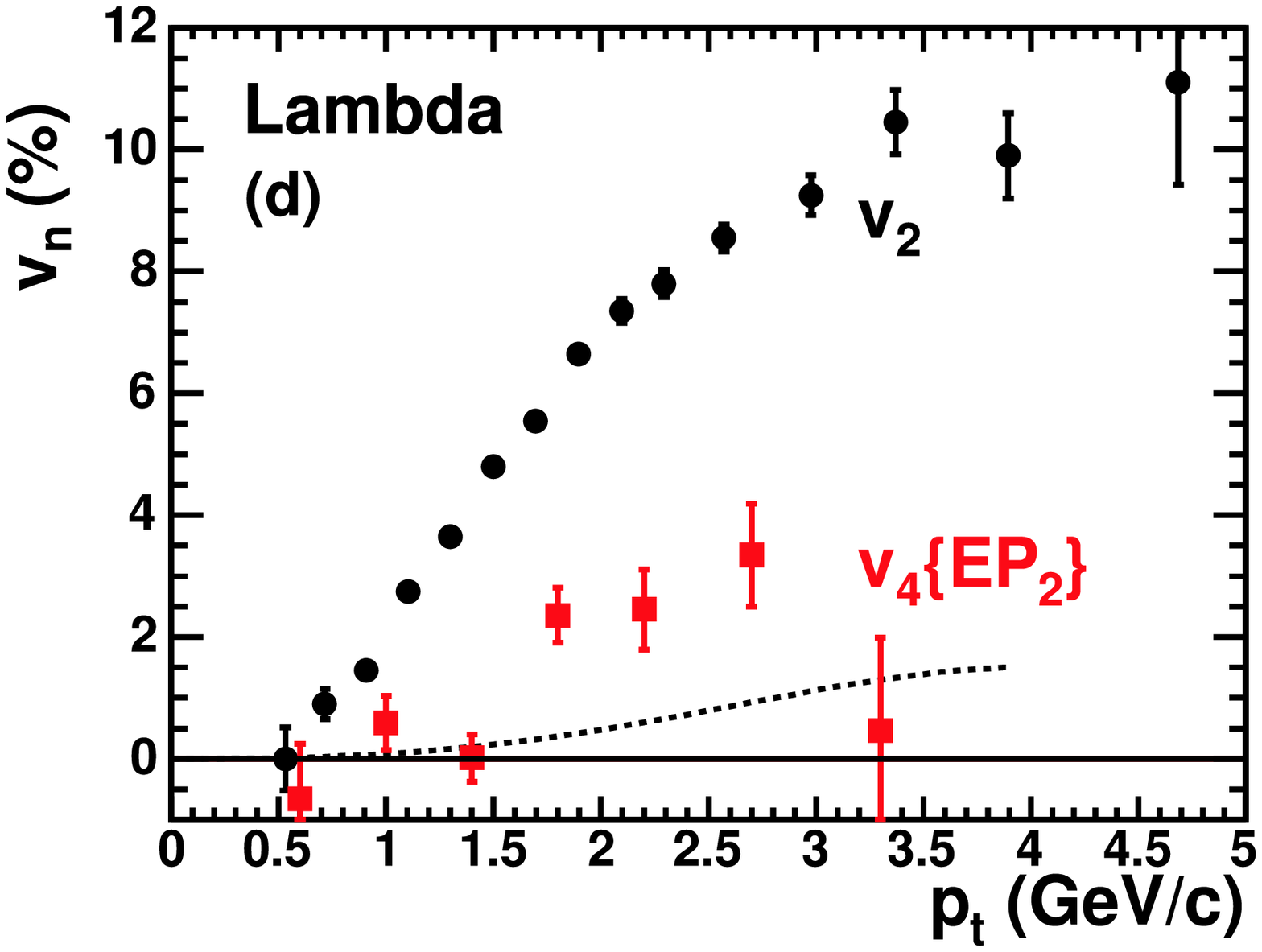}}
  \caption{(color online). $\vF$ and $v_2$ vs.\ $p_t$ for identified
  pions, anti-protons, \ks, and \llam for minimum bias
  collisions. The dashed lines are at $1.2~v_2^2$.
\label{fig:v4PID}}
\end{figure*}

Kolb~\cite{Kolb_v4} pointed out that for large $v_2$ the azimuthal
shape in momentum space described by the $v_n$ Fourier expansion is no
longer elliptic, but becomes ``peanut" shaped. Using our high $p_t$
plateau experimental values, we show this in
Fig.~\ref{fig:polar}. Kolb also gives an equation for the amount of
$v_4$ needed to just eliminate the peanut
waist. Figure~\ref{fig:waist} shows that the experimental $v_4$ values
considerably exceed this value.

Fig.~\ref{fig:v4ptCen} shows the $\vF(p_t)$ values for the individual
centralities with filled elliptic cylinder Blast Wave fits assuming
all charged hadrons have the mass of a pion.

\begin{figure}[!htb]
  \resizebox{\FigFactorSm\textwidth}{!}{\includegraphics{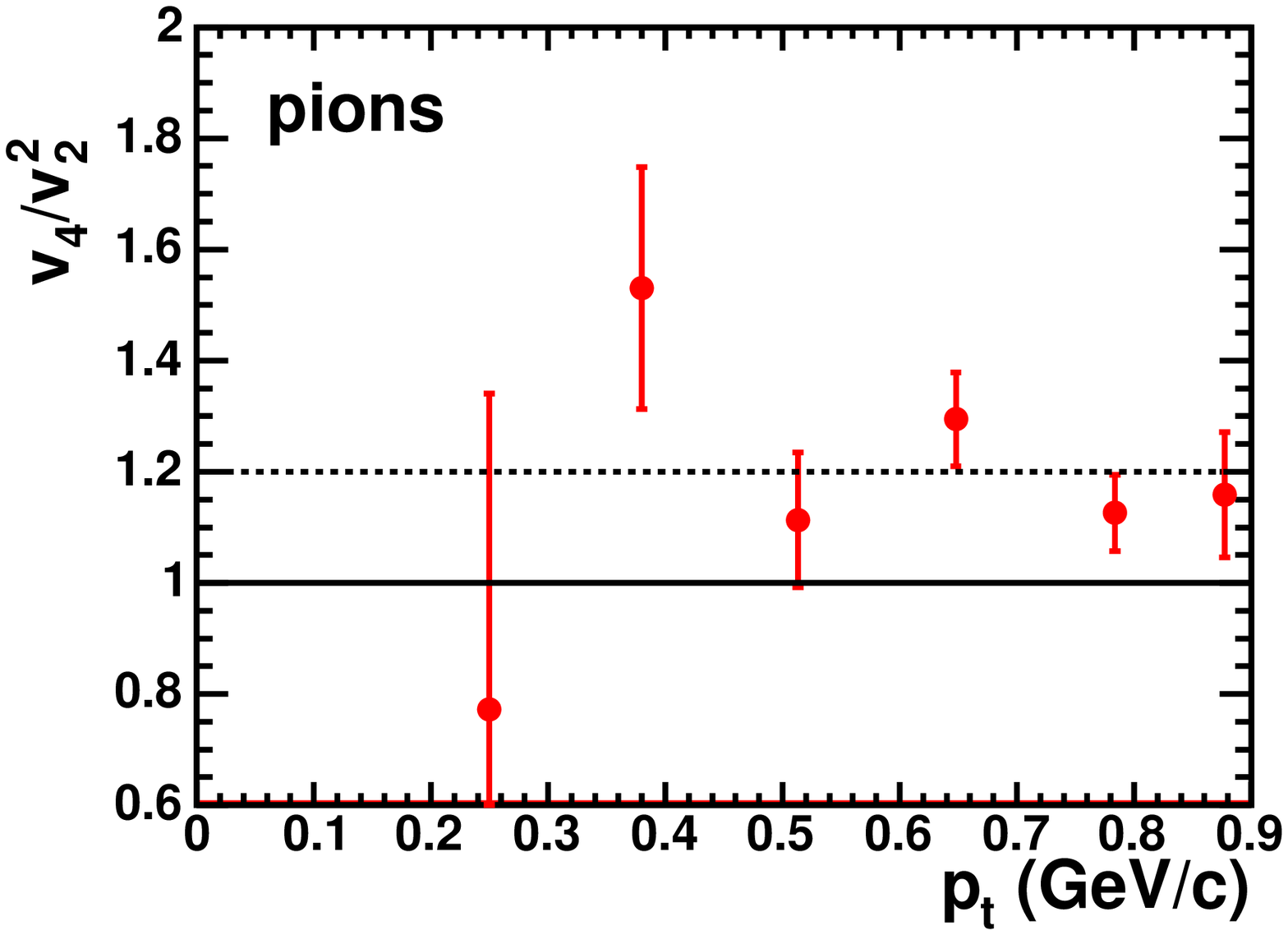}}
  \caption{(color online). The ratio $v_4/v_2^2$ vs.\ $p_t$ for
  identified pions. The dashed line is at $v_4/v_2^2 = 1.2$.
\label{fig:v4PIDratio}}
\end{figure}

Using the probability PID method~\cite{STARPID,TangThesis} for charged
pions and anti-protons, and a topological analysis method for \ks and
\llam, we obtain the $\vF(p_t)$ and $v_2(p_t)$ values shown in
Fig.~\ref{fig:v4PID}. For pions the $v_2^2$ scaling ratio is shown in
Fig.~\ref{fig:v4PIDratio}. To make this graph it was necessary to
combine data points to get reasonable errors bars for the ratio
because the $v_4$ values are so small. The resulting scaling
ratio is consistent with that for charged hadrons shown in the
Fig.~\ref {fig:BW}~(b) ratio graph.

\subsubsection{\label{sec:v4FTPC}The forward regions}       
\begin{figure}[!htb]
  \resizebox{\FigFactor\textwidth}{!}{\includegraphics{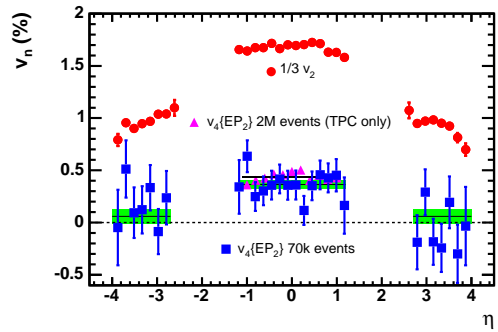}}
  \caption{(color online). Comparison of $v_2$ to \vF\ for
  charged hadrons from minimum bias collisions as a function of
  pseudorapidity. The fourth harmonic (squares) is consistent with
  zero at forward pseudorapidities but not at mid-rapidity. $v_2$
  is shown by circles, scaled by a factor of 1/3 to fit on the
  plot. The larger dataset available for the TPC only (triangles)
  confirms our measurement of \vF\ at mid-rapidities.
  \label{fig:v4Eta}}
\end{figure}

In Fig.~\ref{fig:v4Eta} the fourth harmonic \vF\ shows an average
value of $(0.4\pm0.1)\%$ in the pseudorapidity coverage of the TPC
($|\eta| \lt 1.2$). In contrast, its value of $(0.06\pm0.07)\%$ in the
forward regions is consistent with zero, with a $2\sigma$ upper
limit of 0.2\%. Therefore the relative fall-off of $v_4$ from
$\eta = 0$ to $\eta = 3$ appears to be stronger than for $v_2$. This
behavior is consistent with $v_4\ \propto\ v_2^2$ scaling.


\subsubsection{\label{sec:v4highPt}High $p_t$}       
\begin{figure}[!htb]
  \resizebox{\FigFactorSm\textwidth}{!}{\includegraphics{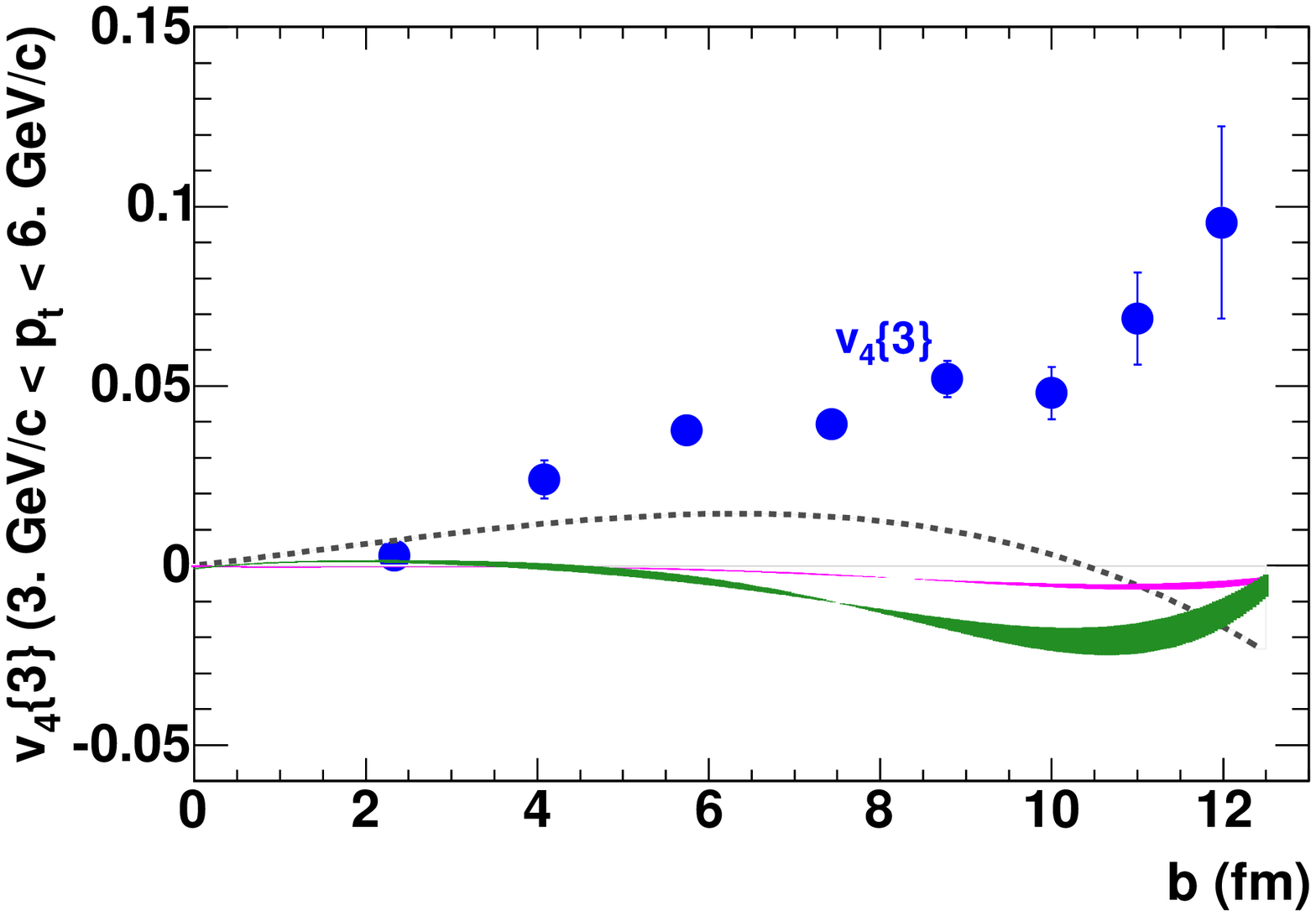}}
  \caption{(color online). High $p_t$ charged hadron $v_4\{3\}$
  integrated for $3 \le p_t \le 6 \ \GeVc$ vs.\ impact parameter
  $b$, compared to models of particle absorption: dashed curve is
  the hard shell, higher narrow band is Woods-Saxon, lower wider band
  is hard sphere. The bands have widths for absorption to match the
  observed range of yield suppression.}
  \label{fig:highPtv4}
\end{figure}

It has been emphasized that $v_4$ has a stronger potential than $v_2$
to constrain jet-quenching model
calculations~\cite{Kolb_v4}. Following the same procedure as described
in Ref.~\cite{STARhighPtV2Corr}, we plot in Fig.~\ref{fig:highPtv4}
the $v_4\{3\}$ from moderately high $p_t$. It should be noted that the
two most peripheral points go up rather than down as they do for
$v_2$, in apparent violation of $v_4/v_2^2$ scaling at this high
$p_t$. We compare the results with the fourth harmonic anisotropy
generated by energy loss in a static medium with a Woods-Saxon density
profile, hard sphere (step function in density), and the extreme case
-- hard shell limit. The results are shown in
Fig.~\ref{fig:highPtv4}. The dashed curve corresponds to the hard
shell; the upper and lower bands corresponds to a parameterization of
jet energy loss where the absorption coefficient is set to match the
suppression of the inclusive hadron yields. The lower and upper
boundaries of the bands around $b=11$ fm correspond to an absorption
that gives suppression factors of 5 and 3, respectively. Note that
compared to the case of $v_2$~\cite{STARhighPtV2Corr}, the
calculations are less sensitive to the suppression factors (narrow
bands). These model calculations cannot reproduce the correct sign of
$v_4$ over the whole range of impact parameters, and neither can they
reproduce the magnitude of $v_4$. A similar observation was made for
the magnitude of $v_2$ in this $p_t$ range in
Ref.~\cite{STARhighPtV2Corr}. In the present case, evidently the
absorption of jet particles is not the dominant mechanism for
producing $v_4$ in this $p_t$ range.


\section{\label{sec:methods}Methods comparisons} In addition to the
standard and scalar product methods already described, there are also
several subevent methods where each particle is correlated with the
event plane of the other subevent. If the subevents are produced
randomly, we will call this the random subs method. If the particles
are sorted according to their pseudorapidity, we will call it the eta
subs method. In these methods, since only half the particles are used
for the event plane, the statistical errors are approximately
$\sqrt{2}$ larger, but autocorrelations do not have to be removed
since the particle of interest is not in the other subevent.

Another method involves fitting the distribution of the lengths of the
flow vectors normalized by the square root of the
multiplicity~\cite{Voloshin-Zhang,OlliQM95,STARcumulants}:
\begin{equation}
  q_n = Q_n/\sqrt{M}       
\label{eq:q}
\end{equation}       
\begin{equation}       
\frac{dP}{q_n d q_n} = \frac{1}{\sigma_n^2}
       e^{\displaystyle{-\frac{v_n^2M+q_n^2}{2\sigma_n^2}}}  
       I_0(\frac{q_n v_n \sqrt{M}}{\sigma_n^2}),
\label{eq:qdist}
\end{equation}       
where $I_0$ is the modified Bessel function and
\begin{equation}       
\sigma_n^2 = 0.5 (1 + g_n).
\label{qWidth}
\end{equation}
Nonflow effects are fit with the parameter $g_n$. The values of
$M$ are in Table~\ref{tbl:centrality}.


\subsection{\label{sec:comaprison}Comparisons}       
\begin{figure}[!htb]
  \resizebox{\FigFactor\textwidth}{!}{\includegraphics{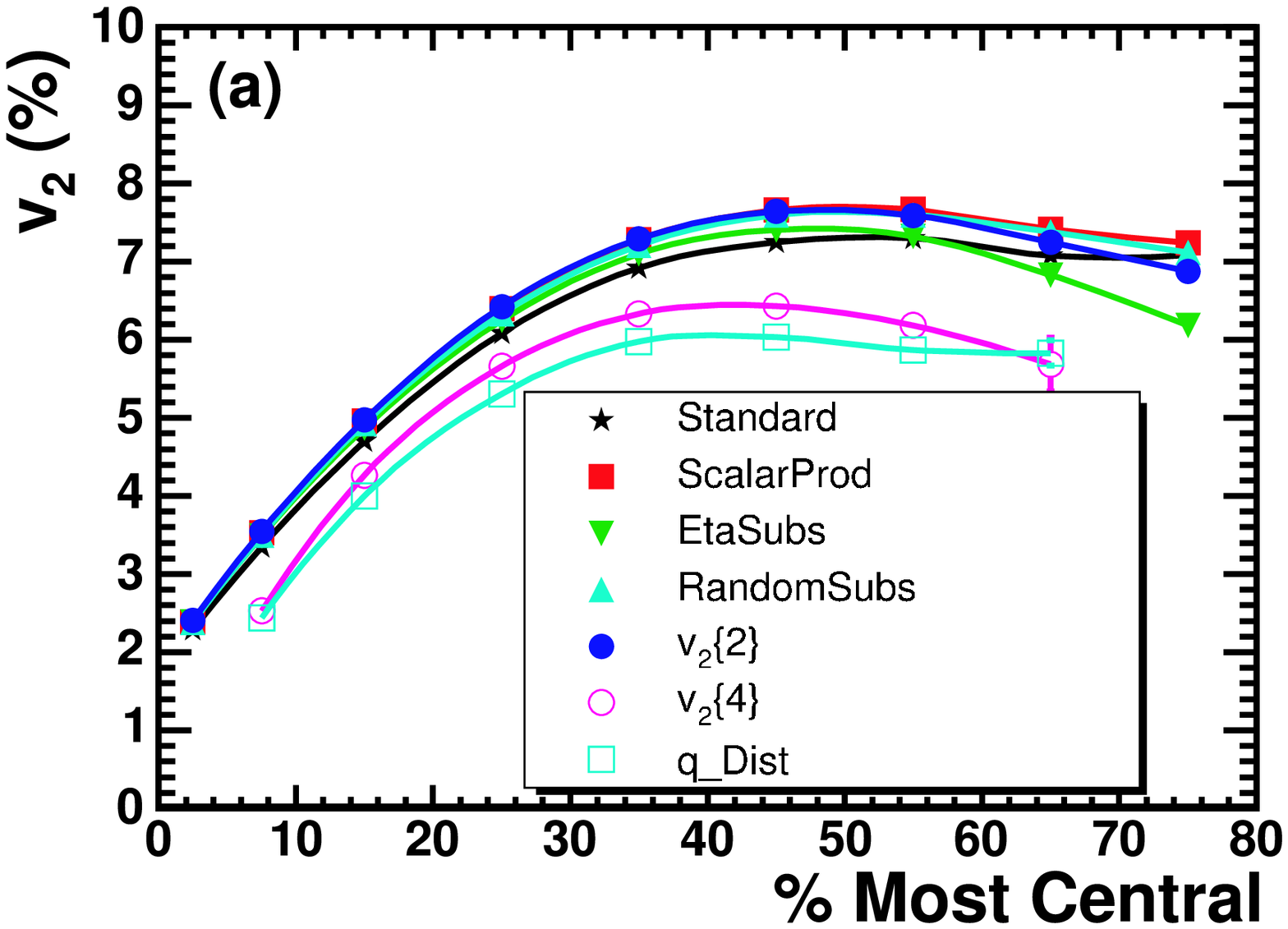}}
  \resizebox{\FigFactor\textwidth}{!}{\includegraphics{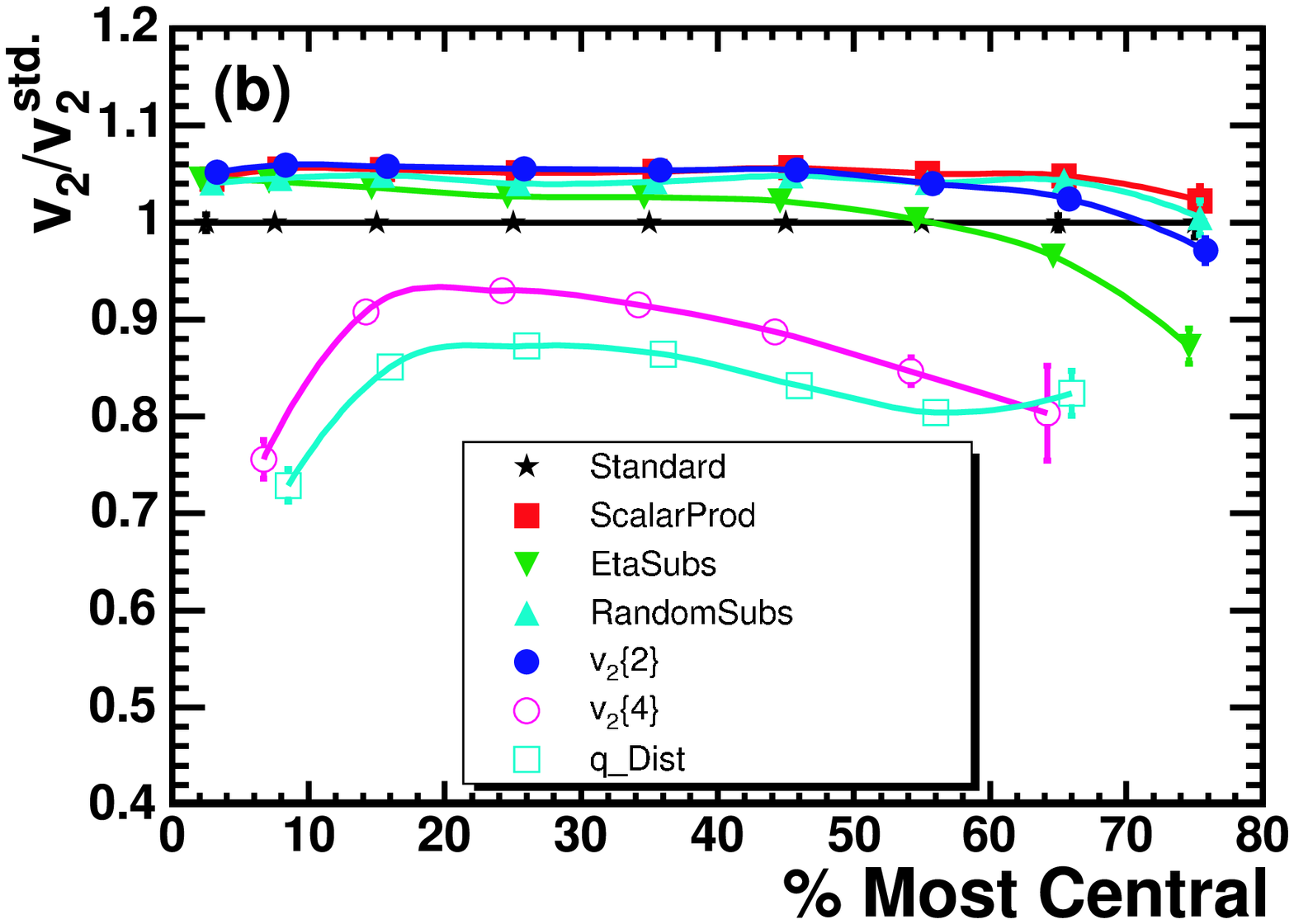}}
  \caption{(color online). Charged hadron $v_2$ integrated over
  $p_t$ and $\eta$ vs.\ centrality for the various methods
  described in the text. In panel (b) is shown the ratio of
  $v_2$ to the standard method $v_2$.
\label{fig:v}}
\end{figure}

To make a precise comparison of the various methods we have calculated
$v_2$ integrated over $p_t$ and $\eta$ for the main TPC, and plotted
it vs.\ centrality in Fig.~\ref{fig:v}~(a). To make the
comparison valid we have used the same events and the same cuts, which
are shown in Table~\ref{tbl:cuts}. The integrated values have not been
corrected for the missing regions beyond the integration limits
given. The systematic error at the lowest $p_t$ values ($\approx$ 0.2
\GeVc) is probably larger than at higher $p_t$, but its contribution
to the integrated $v_n$ values is small because the yield is so low
there.  For constructing the $Q$ vector, linear $p_t$ weighting was
used for all methods except the $q$-distribution method, where no
weighting was used. From the agreement of different software
implementations of the same method, we estimate a relative systematic
error (not included) of at least 2\% of the $v_2$ values shown.

The results fall generally into two bands: those for two-particle
correlations methods, and those for multi-particle methods. The
difference is due either to the decreased sensitivity of the
multi-particle methods to nonflow effects, or to their increased
sensitivity to fluctuation effects~\cite{epsilonMC}. Thus, the
``true'' flow values must be between these two limits. To expand the
graph in order to look for small differences we also have plotted the
ratios to the standard method in Fig.~\ref{fig:v}~(b). It
appears that the standard method is about 5\% lower than the other
two-particle correlation methods. We first thought that this might be
due to nonflow effects affecting the extrapolation in the standard
method from the subevent resolution to the full event
resolution. However, it also could be due to the fact that the
standard method uses twice as many particles as the subevent methods,
and therefore is less sensitive to nonflow effects. But this does not
explain why the scalar product method falls in the band with the
subevents. The values from the eta subevent method decrease for
peripheral collisions. This could be due to decreased nonflow effects
for particles separated in pseudorapidity.


\subsection{\label{sec:nonflow}Nonflow effects and fluctuations}       
\begin{figure}[!htb]
  \resizebox{\FigFactor\textwidth}{!}{\includegraphics{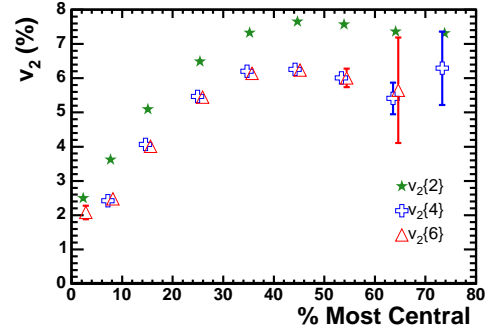}}
  \caption{(color online). Charged hadron $v_2\{2\}$, $v_2\{4\}$, 
  and $v_2\{6\}$ integrated values as a function of centrality.
\label{fig:v246}}
\end{figure} 

Particle correlations which are not correlated with the reaction plane
are called nonflow effects when they affect $v_n$.
Figure~\ref{fig:v246} shows the two, four, and six-particle integral
cumulant $v_2$ values using the cuts in Table~\ref{tbl:cuts}. The
four- and six- particle results agree, showing that nonflow effects
are eliminated already with four-particle correlations.

\begin{figure}[!htb]
  \resizebox{\FigFactor\textwidth}{!}{\includegraphics{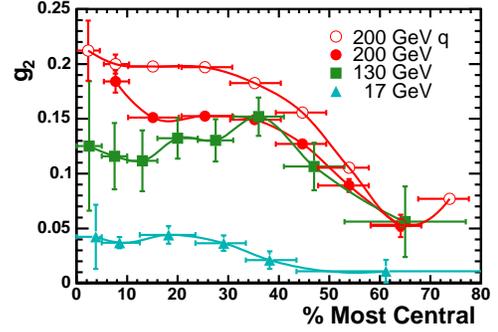}}
  \caption{(color online). The nonflow parameter, $g_2$, as a function
  of centrality. The solid points are from the cumulant method. The
  open circles are from the $q$-distribution method.
\label{fig:nonflow}}
\end{figure}

Nonflow can be calculated by the difference between the squares of the
two-particle and four-particle cumulant $v_2$ values, normalized
with the number of wounded nucleons from
Table~\ref{tbl:centrality}. Thus,~\cite{STARcumulants,NA49,v1{3}}
\begin{equation}
g_2 = N_{WN} \cdot (v_2^2\{2\} - v_2^2\{4\}),
\label{eq:nonflow}
\end{equation}
which is shown in Fig.~\ref{fig:nonflow} for $\sqrtsNN =$ 200 GeV, 130
GeV, and from the SPS at 17.2 GeV~\cite{NA49}. The SPS $g_2$ values
were divided by the multiplicity used and multiplied by $N_{WN}$, both
given in that paper~\cite{NA49}. From the $q$-distribution method of
calculating $v_2$, $g_2$ can be obtained by the increase in the width
of the distribution from Eq.~(\ref{qWidth}). (It should be pointed out
that in these fits, $v_2$ and $g_2$ are somewhat anti-correlated.)
For the $q$-distribution method the $g_2$ values were also divided by
the multiplicity used and multiplied by $N_{WN}$. Thus, all four
results have been renormalized to use the number of wounded nucleons.
Instead of being independent of centrality as originally thought,
$g_2$ seems to decrease somewhat for the more peripheral collisions,
but appears to have the same shape for all the systems. The 17 GeV
results may be different from the others because $g_2$ could vary with
the acceptance of the detector. At 200 GeV it is possible that $g_2$
from the $q$-distribution method is larger than from the cumulant
method because of real fluctuations in $v_2$ broadening the
$q$-distribution. Although the definition of $q$ in Eq.~(\ref{eq:q})
removes most of the multiplicity dependence of $Q$,
Eq.~(\ref{eq:qdist}) still contains the quantity $M$, and thus is
subject to the spread in $M$ in a centrality bin.

\begin{figure}[!htb]
  \resizebox{\FigFactor\textwidth}{!}{\includegraphics{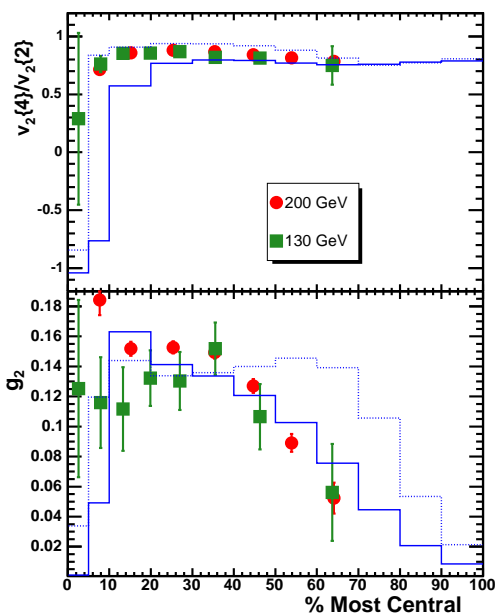}}
  \caption{(color online). Upper panel: The ratio
  $v_{2}\{4\}/v_{2}\{2\}$ for charged hadrons as a function of
  centrality. The lines are a Monte Carlo Glauber model calculation of
  $\eps_{2}\{4\}/\eps_{2}\{2\}$. Lower panel: The nonflow parameter,
  $g_2$, as a function of centrality. The lines are a Monte Carlo
  Glauber model calculation of $N_{WN} \cdot (v_{2} / \eps)^{2}
  \cdot (\eps_{2}^{2}\{2\} - \eps_{2}^{2}\{4\})$. In both panels the
  solid lines assume nucleons, while the dotted lines assume quarks.
\label{fig:nonflowfluct}}
\end{figure}

Fluctuations of the true $v_2$ can lead to an increase in
the $v_{2}\{2\}$ values and an equal decrease in the
$v_{2}\{4\}$ values~\cite{STARcumulants}. In Ref.~\cite{epsilonMC}
initial spatial eccentricity fluctuations are calculated in a Monte
Carlo Glauber (MCG) model and their possible effect on the
determination of elliptic flow is estimated. To do this they take
\begin{eqnarray}
    (\eps\{2\})^2 &=& \langle \eps^2 \rangle \nonumber \\
    (\eps\{4\})^4 &=& 2 \langle \eps^2\rangle^2 - \langle \nonumber
    \eps^4\rangle,
\end{eqnarray}
where the averages are over events and $\eps$ is the eccentricity
which will be defined in Eq.~(\ref{eq:eps}). The physics assumption is
that $v_2 \propto \eps$.  Figure~\ref{fig:nonflowfluct} top panel
shows $\eps\{4\}/\eps\{2\}$ for the quark and nucleon MCG. As with
nonflow, this ratio is smaller than unity over the whole centrality
range, with the largest suppression for the nucleon MCG.  The 130
GeV~\cite{STARcumulants} and 200 GeV data are in between the
calculated values, and are closer to the nucleon (quark) MCG results
for peripheral (central) collisions. When the fluctuations are small
it can be shown that $v_2\{4\} \approx v_2\{6\}$, and from
Fig.~\ref{fig:v246} it is clear that the data indeed support this.

Figure~\ref{fig:nonflowfluct} bottom panel shows the calculated $g_2$
due to eccentricity fluctuations~\cite{epsilonMC}. In contrast to
expectations from nonflow, which would predict a constant value of
$g_2$ vs.\ centrality, the eccentricity fluctuations reproduce the
observed drop of about a factor 3 vs. centrality as observed in the
data.

Thus it appears that either nonflow or fluctuations can explain the
two bands in Fig.~\ref{fig:v}. Most probably it is some of both. Since
nonflow effects and fluctuations raise the two-particle correlation
values, and fluctuations lower the multi-particle correlation values,
the truth must lie between the lower band and the mean of the two
bands. At the moment we can only take the difference of the bands as
an estimate of our systematic error.


\section{\label{sec:models}Model comparisons} This section compares
the experimental results with model calculations. Measurements of
event anisotropy, especially elliptic flow $v_2$, are sensitive to the
early collision
dynamics~\cite{sorge97,sorge99,ollitrault92,shuryak01}. Extracting
physics from the huge set of presented data is done via a variety of
methods, ranging from transport models which include really quite
detailed (and diverse) descriptions of the sub-nuclear dynamics, to
hydrodynamic models which make simplifying assumptions (zero mean free
path and thermalization) rendering all dynamic details irrelevant and
focusing all physics on the equation of state.  We first consider
schematic concepts like coalescence which propose an underlying nature
of the flowing constituents and allow observable tests of scaling
relations implied by those concepts. Finally we use a simple Blast
Wave parameterization, which tries to see whether a consistent picture
of all data can be achieved and to identify what are the required
driving features (like geometric anisotropy at freeze-out, etc).


\subsection{\label{sec:v2Coales}Coalescence of constituent quarks}
Models of hadron formation by coalescence or recombination of
constituent quarks successfully describe hadron production in the
intermediate $p_t$ region ($1.5 \lt p_t \lt 5$~\GeVc)
\cite{Adler:2003kg, STAR_Rcp, reco}. These models predict that at
intermediate $p_t$, $v_2$ will approximately scale with the number of
constituent quarks ($n$) with $v_2/n$ vs.\ $p_t/n$ for all hadrons
falling on a universal curve. When hadron formation is dominated by
coalescence, this universal curve represents the momentum-space
anisotropy of constituent quarks prior to hadron formation. This
simple scaling, however, neglects possible higher harmonics and
possible differences between light and heavy quark flow.

\begin{figure}[!htb]
  \resizebox{\FigFactorBig\textwidth}{!}{\includegraphics{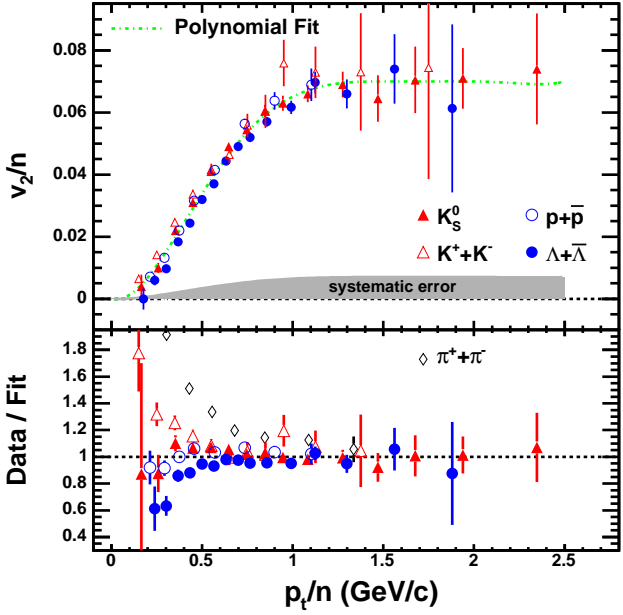}}
  \caption{(color online). Top panel: Identified particle $v_2$ from
  minimum bias collisions. The vertical axis and horizontal axis have
  been scaled by the number of constituent quarks ($n$). Pions are not
  plotted. A polynomial curve is fit to the data. The possible
  systematic error is indicated by the gray band. Bottom panel: The
  ratio of $v_2/n$ to the fitted curve.
\label{fig:v2coales}}
\end{figure}

Figure~\ref{fig:v2coales} (top panel) shows $v_2$ vs.\ $p_t$ for the
identified particle data of Fig.~\ref{fig:v2strange}, where $v_2$ and
$p_t$ have been scaled by the number of constituent quarks ($n$). A
polynomial function has been fit to the shown scaled values. To
investigate the quality of agreement between particle species, the
data from the top panel are scaled by the fitted polynomial function
and plotted in the bottom panel. For $p_t/n \gt 0.6$~\GeVc, the scaled
$v_2$ of \ks, $K^{\pm}$, p+$\mathrm{\overline{p}}$, and \llam lie on a
universal curve within statistical errors. The pion points, however,
deviate significantly from this curve even above 0.6 \GeVc. This
deviation may be caused by the contribution of pions from resonance
decays~\cite{decayv2}. Alternatively, it may reflect the difficulty of
a constituent-quark-coalescence model to describe the production of
pions whose masses are significantly smaller than the assumed
constituent-quark masses~\cite{reco}.

At the end of Sec.~\ref{sec:nonflow} we estimated that the $v_2$
values from two-particle correlations could be systematically high by
between about 10 to 20\%. This was based on the integrated values for
charged particles and we do not know yet how this varies with $p_t$
and particle type. However, to indicate this estimated systematic
error a shaded band of 10\% is shown in Fig.~\ref{fig:v2coales} (top
panel).

\begin{figure}[!htb]
  \resizebox{\FigFactorBig\textwidth}{!}{\includegraphics{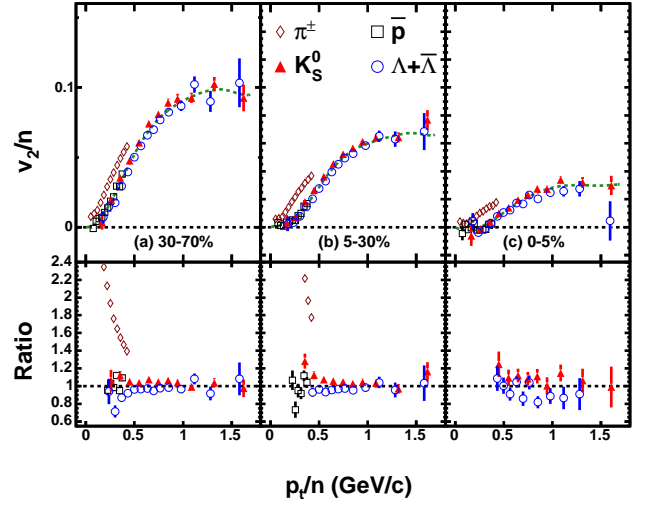}}
  \caption{(color online). Top panels: The $v_2$ of $\pi^{\pm}$,
  $\overline{p}$, \ks, and \llam from three centrality bins
  (30--70\%, 5--30\%, and 0--5\% of the collision cross section)
  scaled by the number of constituent quarks ($n$) vs.\
  $p_t/n$. Polynomial curves are fit to the data excluding the
  pions. Bottom panel: The ratios of $v_2/n$ to the fitted curves.
\label{fig:v2coalesCen}} 
\end{figure}

The $v_2/n$ of $\pi^{\pm}$, $\overline{p}$, \ks, and \llam from three
centrality intervals are shown in the top panels of
Fig.~\ref{fig:v2coalesCen}. The \ks and \llam values are from
Ref.~\cite{STAR_Rcp}. In the bottom panels, the ratios to the fitted
curves are shown. The most central data (0--5\%) are thought to be
affected by nonflow correlations (See Sec.~\ref{sec:methods}). For the
30--70\% and 5--30\% centrality intervals, the $v_2$ of
$\overline{p}$, \ks and \llam agree with constituent-quark-number
scaling for the expected $p_t/n$ range above 0.6~\GeVc\ to within
10\%.

Figure~\ref{fig:v2strange} showed that the data for the heavier
baryons seem to cross over the data for the mesons at sufficiently
high $p_t$. The data in Fig.~\ref{fig:v2RICH} are consistent with
this. In the low $p_t$ region the heavier particles have lower
$v_2$ values as expected for the mass ordering from hydrodynamics. In
the intermediate $p_t$ coalescence plateau region the three quark
baryons have a larger $v_2$ than the two quark mesons. Thus the
experimentally observed cross-over is thought to be due to a change in
the particle production mechanism.

\begin{table}[hbt]
\caption{The ratio $v_4/v_2^2$ for all $p_t$ and only for $p_t/n > 0.6$ \GeVc.} 
\begin{center}
\begin{tabular}{l c c} \hline \hline
       & all $p_t$	& $p_t/n > 0.6$ \GeVc  \\ \hline   
  $h^{\pm}$   & 1.17 $\pm$ 0.01	& 1.14 $\pm$ 0.02  \\   
  $\pi^{\pm}$  & 1.19 $\pm$ 0.04 &   \\   
  \ks  & 3.1 $\pm$ 1.2	& 2.5 $\pm$ 1.3	\\   
  $\overline{p}$   & 1.46 $\pm$ 0.53	&  \\   
  \llam   & 0.97 $\pm$ 0.18	& 0.87 $\pm$ 0.22   \\   
\hline \hline
\end{tabular}   
\end{center}   
\label{tbl:v4v22}   
\end{table}   

From a simple parton coalescence model one can
calculate~\cite{coalescence} the observed $v_4/v_2^2$ scaling
ratio in terms of the same quantity for the quarks. The relationships
between meson ($M$) or baryon ($B$) $v_4/v_2^2$ and quark ($q$)
$v_4/v_2^2$ are
\begin{equation}
 \left[v_4/v_2^2 \right]^M_{p_t} \approx 1/4 + (1/2) \left[v_4/v_2^2 \right]^q_{p_t/2},
\label{eq:mesons}
\end{equation}
and 
\begin{equation}
 \left[v_4/v_2^2 \right]^B_{p_t} \approx 1/3 + (1/3) \left[v_4/v_2^2 \right]^q_{p_t/3}.
\label{eq:baryons}
\end{equation}
These can be rearranged~\cite{coalescence} to relate $v_4/v_2^2$ for
mesons and baryons:
\begin{equation}
  \left[v_4/v_2^2 \right]^B_{p_t/3} \approx (2/3) \left[ v_4/v_2^2 \right]^M_{p_t/2} + 1/6.
\label{eq:baryonsFromMesons}
\end{equation}
The observed $v_4/v_2^2$ scaling ratios, which appear to be fairly
independent of $p_t$ in Figs.~\ref{fig:BW}~(b), \ref{fig:v4PID} and
\ref{fig:v4PIDratio}, are shown in Table~\ref{tbl:v4v22}. Although in
Fig.~\ref{fig:v2coales}, quark-number scaling is shown to work within
errors at $p_t/n > 0.6$~\GeVc\ for all particles except pions, it
appears that $v_4/v_2^2$ scaling may be applicable over a wider range
of $p_t$. Charged hadrons are in the Table but should be used with
care because they represent a complicated superposition of baryons and
mesons from different values of $p_t/n$ where the B/M ratio is
strongly dependent on centrality and we cannot even assume that the
values are a good estimator for mesons. The kaon values are not
accurate enough to test the above equation.  Even though the pions are
known to deviate from the constituent quark number coalescence
predictions, we can calculate with Eq.~\ref{eq:baryonsFromMesons},
from the charged pions for the wide $p_t$ range, that $v_4/v_2^2$ for
baryons should be 0.96 $\pm$ 0.03. This is compatible with the values
for anti-protons and \llam in
Table~\ref{tbl:v4v22}. Equation~\ref{eq:baryonsFromMesons} would be
valuable for testing the concept of quark coalescence in an
equilibrated medium, but the accuracy of the data so far do not allow
a conclusion.

If, in addition, one assumes~\cite{partonScaling,coalescence} that
the scaling relation for the partons is
\begin{equation}
  v_4^q = (v_2^q)^2,
\label{eq:partonScaling}
\end{equation}
then from Eq.~(\ref{eq:mesons}) $v_4 / v_2^2 = 1/4 + 1/2 = 3/4$.  For
baryons this ratio from Eq.~(\ref{eq:baryons}) is $1/3 + 1/3 = 2/3$,
which is even smaller.  But, one can see in Table~\ref{tbl:v4v22} that
experimentally this ratio is close to 1.2 for charged hadrons and
pions, so that either the parton scaling relation
(Eq.~(\ref{eq:partonScaling})) must have a proportionality constant of
about 2, or the simple coalescence model needs improvement.


\subsection{\label{sec:trans}Transport models} Most of the transport
model analyses were done for charged hadrons, but we will only compare
some of the models with identified hadrons. Microscopic hadronic
transport calculations under-predict the absolute amplitude of $v_2$
by a factor of 2 to 3. However, most of the observed features, like
mass hierarchies in both the low $p_t$ region and the meson-baryon
order, are seen in hadron transport model
calculations~\cite{bleicher02}. The strength of $v_2$ should be
sensitive to the density and interaction frequency of the
constituents. Indeed, when reducing the hadron formation time, the
$v_2$ values are found to increase~\cite{bleicher02}. In addition, the
tests with the parton cascade models AMPT~\cite{zwlin02} and
ZPC~\cite{zpc99} give the correct mass hierarchy but require a large
parton cross section in order to mimic the early development of $v_2$
at mid-rapidity. In ultra-relativistic nuclear collisions, hadrons may
not be the right degrees of freedom to describe the early dynamics. At
large values of pseudorapidity, however, the AMPT~\cite{Ko04} model
seems able to describe the $v_1$, $v_2$, and $v_4$ results without the
large parton cross sections and string melting. At all
pseudorapidities, at the later stage, when particle density becomes
dilute, transport effects will become
important~\cite{teaney01,bravina04}.

For $v_4$ the parton cascade model AMPT~\cite{partonScaling} with
string melting and a large parton cross section, does calculate
reasonable values. However, the calculated proportionality constant in
Eq.~(\ref{eq:partonScaling}) is about 1, while our data with a simple
coalescence model~\cite{reco} imply it to be about 2.


\subsection{\label{sec:hydro}Hydrodynamic models} Azimuthal momentum
anisotropies in the final state are generated by particle
re-interactions from azimuthal spatial anisotropy in the initial
state. In the hydrodynamic framework, these re-interactions are
modeled by assuming zero mean free-path and therefore local
thermalization. Hydrodynamic calculations have been successful at
reproducing previously published data on $v_2$ and
spectra~\cite{heinz04,kolb04,heinz04a}.

\begin{figure} [!htb]
  \resizebox{\FigFactorBig\textwidth}{!}{\includegraphics{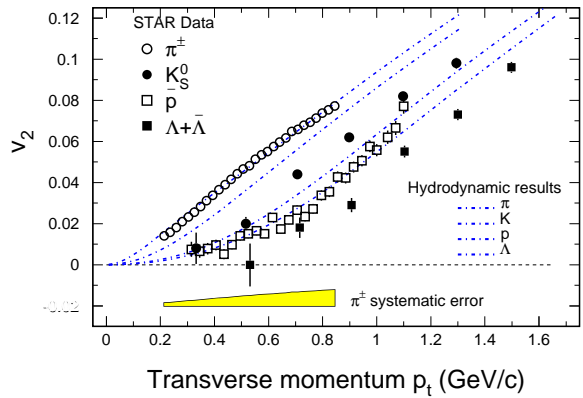}}
  \caption{(color online). $v_2(p_t)$ for charged $\pi$, \ks,
  $\overline{p}$ and \llam from minimum bias collisions. Hydrodynamic
  calculations~\cite{pasi01,pasi03} are shown as dot-dashed
  lines. The possible systematic error is shown at the bottom.}
\label{fig:hydro_mbv2} 
\end{figure}

Hydrodynamic calculations have been shown in Figs.~\ref{fig:v2PID},
\ref{fig:v2RICH}, \ref{fig:kaon_v2}, and \ref{fig:v2strange}, with
reasonable agreement with the $v_2$ and $v_2\{4\}$ data up to $p_t$ of
1-2 \GeVc. Additional results for $v_2$ at low $p_t$ from minimum bias
collisions, are shown in Fig.~\ref{fig:hydro_mbv2}. Results of \ks and
\llam are from Ref.~\cite{STAR_Rcp}. The hydrodynamic
calculations~\cite{pasi01,heinz04,pasi03} are consistent with the
experimental results considering the systematic errors, such as the
matching of the centralities are not included. Also, as described in
Sec.~\ref{sec:methods}, the data could be 10 to 20\%
systematically high. To indicate this in the plot a band of 10\%
of the charged pions is shown. The characteristic hadron mass
ordering of $v_2$ is seen in the low $p_t$ region, where at a given
$p_t$, the higher the hadron mass the lower the value of $v_2$. This
supports the hypothesis of early development of collectivity and
possible thermalization in collisions at RHIC~\cite{heinz04,heinz04a},
although the underlying mechanism for the equilibration process
remains an open issue.

As seen in Fig.~\ref{fig:v2coales} the observed values of $v_2$
saturate and the level of the saturation seems dependent on the number
of constituent quarks (n) in the hadron. The saturation value is about
$0.07 n$ for $p_t/n >$ 1 \GeVc. Hydrodynamic calculations do not
saturate in this $p_t$ region.

\begin{figure} [!htb]
  \resizebox{\FigFactorBig\textwidth}{!}{\includegraphics{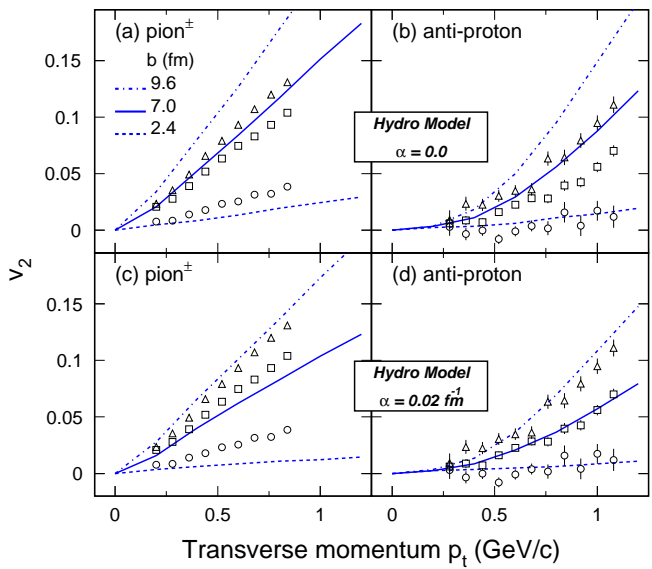}}
  \caption{(color online). Charged $\pi$ plots on the left and
  $\overline{p}$ plots on the right for $v_2$ for three centrality
  bins are shown as a function of $p_t$. The data are from
  centralities 40--50\% (open triangles), 20--30\% (open squares) and
  0--5\% (open circles).  The corresponding results of a hydrodynamic
  calculation are shown as dot-dashed lines, solid-lines, and
  dashed-lines, respectively. Plots on the top are for $\alpha = 0$
  and plots on the bottom are for $\alpha = 0.02 \ \rm{fm}^{-1}$. Here
  $\alpha$ determines the initial velocity kick for the hydrodynamic
  model calculation~\cite{kolb02}.}
\label{fig:hydro_v2} 
\end{figure}

Figure~\ref{fig:hydro_v2} shows the centrality dependence of pion and
anti-proton $v_2$ compared with hydrodynamic
results~\cite{kolb02}. The three centrality bins shown are described
in Table~\ref{tbl:centrality}. Systematic uncertainties, such as the
matching of the centralities, are not included. Also, from the
Fig.~\ref{fig:v}~(b) ratio graph in Sec.~\ref{sec:methods} it
can be seen that the 0--5\% centrality data could be 25\% high. An
important concern for the 0--5\% centrality bin is the
fluctuations. Just averaging over the spread in impact parameters in
this bin could lower $v_2$ a factor of
two~\cite{STARcumulants,epsilonMC}. In the hydrodynamic calculation,
the decoupling temperature was set to 100 MeV. In order to fit the
$p_t$ spectra of (anti-)protons, the hydrodynamic evolution was
started with an initial transverse velocity kick of $\tanh(\alpha
\cdot r)$, where $\alpha$ is a parameter~\cite{kolb02}. The results
for $v_2$ are shown in Fig.~\ref{fig:hydro_v2}. For $\alpha = 0$,
Fig.~\ref{fig:hydro_v2} (a) and (b), neither pion nor anti-proton
results can be fitted . For $\alpha =0.02 \ {\rm fm}^{-1}$,
anti-proton (d) $v_2$ can be fitted reasonably well but, for pions
(c), the model results still miss the data. It appears that with the
initial velocity, there is too much kick for pions at both mid-central
and central collisions.  Due to their light mass, perhaps pions
decouple from the system relatively earlier than protons, as also
indicated in the pion interferometry results~\cite{rhic130hbt}. It
seems that for the 40--50\% centrality data the hydro calculations
over-predict the data, which is not surprising for peripheral
collisions.

Both Hirano~\cite{hirano} and Heinz and Kolb~\cite{noTherm} explain
the fall-off of $v_2$ at high $\eta$ as being due to incomplete
thermalization. The particle density, $dN/dy$, also falls off in the
same way, and at high $\eta$ is similar to that at mid-rapidity at the
SPS~\cite{NA49}, where the flow values are also lower. Possibly, the
lower particle density leads to less thermalization, and therefore
smaller $v_2$ values.

Hydrodynamic inspired fits have been done for
spectra~\cite{heinz93}. Csan\'{a}d {\it et al.} now report results
where the authors claim that the resulting $p_t$ spectra,
interferometry parameters, and anisotropy can all be
fitted~\cite{csanad04}. In particular, they have a fall-off of $v_2$
at high $\eta$. But their $v_1(\eta)$ has a large wiggle near
mid-rapidity which is not observed. They further determined the source
parameters and concluded that about 15\% of the hadrons are emitted
directly from the super-heated region.

So far there have been very few model calculations of $v_4$. However,
the magnitude and even the sign of $v_4$ are more sensitive than $v_2$
to initial conditions in the hydrodynamic calculations~\cite{Kolb_v4}.
This calculation predicted $v_4/v_2^2$ to vary from 0.7 to 0.3 going
from low to high $p_t$, which is about a factor of two lower than
observed in the Fig.~\ref{fig:BW}~(b) ratio graph and
Table~\ref{tbl:v4v22}. This calculation also predicted a strongly
negative $v_6$, which is not observed~\cite{STARv1v4}.


\subsection{\label{sec:BW}Blast Wave models} Blast Wave models
parameterize the coordinate and momentum freeze-out configuration
generated in hydrodynamic calculations. In a self-consistent
hydrodynamic calculation, this configuration is determined by the
Equation of State and freeze-out prescription; in Blast Wave
calculations, parameters of the distribution may be varied arbitrarily
to fit the data.  In this sense, Blast Wave is a ``toy'' model useful
mainly to characterize the data and determine the magnitude of thermal
(random) motion, collective motion, geometry, etc. The model also
provides parameters that can be used to study the evolution of flow
varying the initial conditions, which in this paper is achieved by
varying centrality.

The present paper uses two versions of the Blast Wave
parameterization. In the first one, all particles are emitted from a
surface shell boosted by a constant flow
velocity~\cite{STARPID,STARv4}. In the second one, particles are
emitted from a filled elliptic cylinder boosted perpendicular to the
surface of the cylinder, and with a linear transverse rapidity profile
inside the cylinder~\cite{BlastWave}. In this paper, unless otherwise
specified, Blast Wave fits have referred to the filled elliptic
cylinder version.

\begin{figure} [!htb]
  \resizebox{\FigFactor\textwidth}{!}{\includegraphics{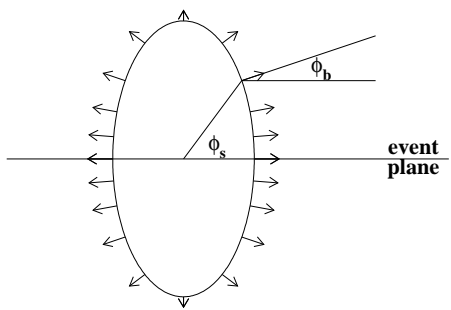}}
  \caption{(color online). Schematic illustration of an
  elliptical sub-shell of the source. Here, the source is extended in
  the direction out of the event plane ($R_y > R_x$). Arrows represent
  the direction and magnitude of the flow boost. In this example,
  $\rho_2 > 0$. From Ref.~\cite{BlastWave}.}
\label{fig:ellipse} 
\end{figure}

In recent versions of Blast Wave models, the system is assumed
boost-invariant in the beam direction.  As suggested in
Fig.~\ref{fig:ellipse} for the filled elliptic cylinder, the geometry
in the transverse direction is a filled ellipse with the major axis
aligned with the reaction plane or perpendicular to it.  One may
quantify the geometrical anisotropy of the system with the
eccentricity
\begin{equation}
\eps \equiv \frac{R^2_y-R^2_x}{R^2_y+R^2_x} ,
\label{eq:eps}
\end{equation}
where the $x$ direction is in the reaction plane.  Superimposed on a
randomly-directed energy component quantified by a temperature, $T$,
each geometrical cell of the system is boosted ``outward'' by a
velocity (flow) field.  Here, ``outward'' indicates the direction
normal to the surface of the elliptical shell on which the element
sits.  The magnitude of the flow field vanishes (by symmetry) at the
center of the system and grows linearly with the distance from the
center, reaching its maximum at the transverse edge of the system
(here assumed to be a sharp, non-diffuse edge).  The average value of
the flow magnitude is quantified by a parameter $\rho_0$. The flow
magnitude may be larger (or smaller) for sources emitting in the $x$-
versus the $y$-direction; the magnitude of this boost oscillation with
azimuthal angle is quantified by the parameters $\rho_2$ and $\rho_4$.
In Fig.~\ref{fig:ellipse}, a larger in-plane than out-of-plane boost
(corresponding to $\rho_2 > 0$) is suggested by the longer boost
angles in-plane.

Several parameters of the system affect $v_2$. Obviously, the larger
the magnitude of $\rho_2$, the larger the momentum-space anisotropy.
Further, the geometric anisotropy plays a role even if the boost
strength is identical in all directions ($\rho_2=\rho_4=0$), if
$R_y>R_x$ ($R_y<R_x$) it is clear from Fig.~\ref{fig:ellipse} that a
greater (lesser) number of elements boost particles into the reaction
plane, resulting in anisotropy in azimuthal momentum space.  Finally,
it is clear that the temperature, $T$, plays a role, since if the
random energy component is dominant ($T$ larger than the rest mass),
momentum anisotropies will be reduced.  An extensive discussion of the
interplay between these effects may be found in Ref.~\cite{BlastWave}.

To summarize, the free parameters of the fits in the shell case are
$T$, $\rho_0$, $\rho_2$, $\rho_4$, $s_2$ and $s_4$, where $T$ is the
temperature parameter, the $\rho_n$ are the harmonic coefficients of
the source element boost in transverse rapidity, and the $s_n$ are the
harmonic coefficients of the source density which boosts into a
particular direction. In previous parameterizations~\cite{STARPID}
where there was no $\rho_4$, $\rho_2$ was called $\rho_a$. In the
filled ellipse case the free parameters are $T$, $\rho_0$, $\rho_2$,
$\rho_4$, $R_x$, and $R_y$, where $R_x$ is the in-plane radius of the
ellipse and $R_y$ is kept constant at a non-zero value. In fitting
data with a surface shell model $\rho_0$ is about $2/3$ as large as
for a solid cylinder with a linear profile. The eccentricity is
approximately equal to $2 s_2$. For an ellipse, the parameter $s_4$ is
approximately equal to $s_2^2$. The actual equations used are given in
Appendix~\ref{sec:BWapp}.

First we verified that the hydrodynamic calculations reported in
Ref.~\cite{Kolb_v4} can be successfully fit by the Blast Wave model
with reasonable parameters: $T$ = 93 MeV, $\rho_0$ = 0.91, $\rho_2$ =
0.080, $\rho_4$ = 0.0017, $\eps$ = 0.122. Since the hydro had no error
bars there is no \chisqrndf. While spectra and $v_2$ are well
reproduced up to $p_t = $ 1.5--2 \GeVc, the $p_t$ dependence of $v_4$
appears quadratic in the Blast Wave, while rather linear in the
hydrodynamic calculation.

We have seen in Fig.~\ref{fig:v2Pions} that the Blast Wave
parameterization does a good job at simultaneously reproducing pion,
kaon and anti-proton $v_2$. The fits are performed simultaneously to
spectra as well as on $v_2$ and $v_4$, in order to be
over-constrained. Pion, kaon and proton spectra (not shown) are well
reproduced. Because spectra have typically more data points and
smaller error bars, both $T$ and $\rho_0$ can be determined, while
$\rho_2$, $\rho_4$, and $\eps$ are constrained by the $v_n$. The total
$\chi^2$ per degree of freedom varies for different centralities
around an average value of 56/65, without exhibiting any specific
dependences. The average $\chi^2$ per data point is 14/6 for pions,
7/4 for kaons and 17/10 for protons. When looking at individual data
sets (e.g pion $v_2$, proton spectra), the $\chi^2$ is compared to the
number of data points because the degrees of freedom can only be
calculated including all the data points as each parameter is
constrained by more than one data set.  Because the $v_2$ error bars
are small (less than 5\%) compared to the spectra error bars (between
5 and 10\%) the total $\chi^2$ is dominated by the contributions from
the $v_2$ results. The calculation fits the peculiar negative values
of the anti-proton $v_2$ in Fig~\ref{fig:v2Pions}~(c) in central
collisions with $p_t$ below 0.5 \GeVc. This feature is reproduced when
$\rho_2$ is significant while the thermal velocity is small. In this
case the flow boost is strong enough that it suppresses the low $p_t$
anti-proton emission in-plane compared to
out-of-plane~\cite{pasi01}. When the eccentricity is sufficiently
large this phenomenon does not take place. The pion $v_2(p_t)$ data
points in Fig~\ref{fig:v2Pions}~(a) are similar in the three
most peripheral bins. However, the anti-proton values are not, and
thus meaningful fits are still possible. The $p_t$ ranges in \GeVc\
used for the Blast Wave fits where the data had reasonable error bars
were 0.4 to 1.0 for pions, 0.15 to 0.5 for kaons, and 0.3 to 1.1 for
anti-protons.

\begin{figure}[!htb]
  \resizebox{\FigFactor\textwidth}{!}{\includegraphics{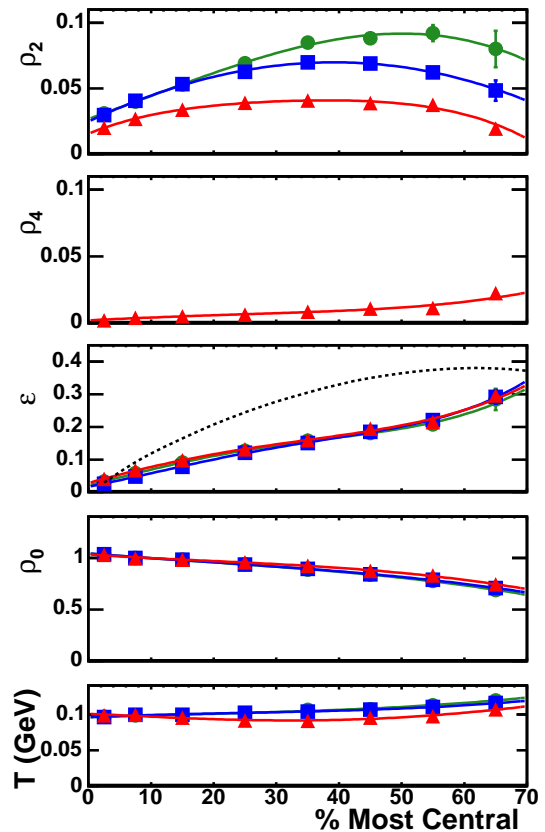}}
  \caption{(color online). The Blast Wave parameters $\rho_2$,
  $\rho_4$, $\eps$, $\rho_0$ and T plotted vs.\ centrality. The
  circles are for pions from a two-particle cumulant analysis, the
  squares for pions from a standard event-plane analysis, and the the
  triangles for charged hadrons from a standard analysis. The lines
  are polynomial fits. In the middle panel the initial geometrical
  eccentricity is also plotted as a dashed line. The actual parameter
  values are available at
  \protect\url{http://www.star.bnl.gov/central/publications/}~.  
  \label{fig:BWpars}}
\end{figure}

The Blast Wave parameters obtained from fitting $v_2$ and $v_4$ data
are shown in Fig.~\ref{fig:BWpars}. They provide a good way to
systematize a large amount of experimental data. It should be
emphasized that other formulations of the Blast Wave model would give
different fit parameters~\cite{tomasik}. As the parameters $T$
and $\rho_0$ are constrained mostly by spectra, they agree with the
values published~\cite{STARspectra200}. $\rho_2$ and $\eps$ are fully
constrained by the $v_2$ data. $\rho_2$ reaches a maximum in the
centrality region 30--60\%. This is easily understood recalling that
in this centrality region, the initial spatial azimuthal anisotropy of
the system is large, while the initial energy density is still large
enough to trigger a significant collective expansion. This expansion
is clearly visible comparing the initial and final eccentricities. The
system spatial deformation is a maximum in the region where the
azimuthal push quantified by $\rho_2$ is a maximum. Thus, the Blast
Wave parameterization provides an intuitive self-consistent
description of the data.

\begin{figure}[!htb]
  \resizebox{\FigFactor\textwidth}{!}{\includegraphics{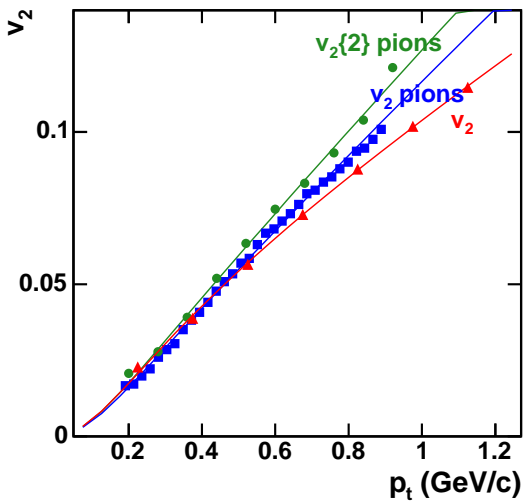}}
  \caption{(color online). For centrality 20--30\% we show $v_2\{2\}$
  for pions (circles), $v_2$ for pions (squares), and $v_2$ for
  charged hadrons (triangles). The solid lines are Blast Wave fits.
\label{fig:v2Comp}}
\end{figure}

For one centrality we show in Fig.~\ref{fig:v2Comp} the charged hadron
results from this standard event-plane analysis, together with pion
results for a standard analysis and a two-particle cumulant
analysis. As shown in Sec.~\ref{sec:methods}, the integrated
two-particle cumulant $v_2\{2\}$ values are usually 5\% higher than
the standard $v_2$ values. The charged hadron values are somewhat
smaller than the pion values, because of the presence of protons. Even
though the flow values are fairly close, the $\rho_2$ fit parameters
shown in Fig.~\ref{fig:BWpars} differ appreciably. This is because the
$\eps$ values come out the same and the small differences in the $v_2$
values are all forced into the $\rho_2$ values. It appears that the
$\eps$ values are at least half as large as the initial eccentricities
of the overlap region.

Both hydrodynamic calculations and Blast Wave fits can well reproduce
transverse momentum spectra and second-harmonic anisotropy ($v_2$).
However, as mentioned above, hydrodynamic calculations do not agree
with measured values of $v_4$.  The question, then, is whether Blast
Wave parameters may be adjusted to simultaneously fit $v_2$ and $v_4$,
hopefully providing useful feedback to theorists doing the
hydrodynamic calculations. Blast Wave fits to $v_4$ are shown in
Fig.~\ref{fig:v4ptCen}.

Even with only second-harmonic anisotropies in flow strength and
spatial geometry, fourth-harmonic momentum-space anisotropies ($v_4$)
are produced in Blast Wave calculations.  Thus it is possible that one
could generate $v_4$ without any fourth harmonic anisotropy $\rho_4$
in Eq.~(\ref{eq:rho})~\cite{coalescence}. Using the surface shell
Blast Wave model, we have fit the $v_2$ data using only $\rho_2$ and
$s_2$, and then calculated $v_4$ as shown in Fig.~\ref{fig:BW} as the
dashed lines. The calculated $v_4$ values are much too small,
indicating that a real fourth harmonic term is necessary. Then we
allowed $\rho_4$ and $s_4$ to vary as well and obtained the fits shown
in Fig.~\ref{fig:BW} as solid lines. The fact that these parameters
are significant suggests that the spatial distribution of the system
initial state has a significant fourth harmonic component, which
translates into a fourth harmonic flow oscillation.


\section{\label{sec:concl}Conclusions} All the presently available
STAR data for anisotropic flow in Au+Au collisions at $\sqrtsNN = 200$
GeV are presented for charged particles and for identified
species. Agreement between flow data for STAR and other RHIC
experiments is good. New evidence confirms our earlier finding that
elliptic flow is in-plane at RHIC. $v_2$ as a function of
pseudorapidity is not flat, but confirmed to be bell shaped. A
detailed comparison of flow analysis methods, shows that either
nonflow effects or fluctuations can explain the difference between
$v_2$ from two-particle correlation results and multi-particle
correlation results. The mass dependence of $v_2$ at low $p_t$ follows
the pattern predicted by hydrodynamic models, but a transition to a
behavior consistent with quark coalescence at higher $p_t$ is
observed. For identified particles, $v_2$ scales with the number of
constituent quarks, $n$, within errors above $p_t/n \sim 0.6$ \GeVc\
for charged and neutral kaons, for anti-protons, and for \llam\
hyperons. This supports the picture of hadron production via
coalescence of constituent quarks involved in collective anisotropic
motion. If confirmed it would be a strong argument for the
deconfinement reached in the system. Only pions deviate from this
behavior, which partially can be explained by the large resonance
decay contribution to pion production, and by the light pion mass. For
the higher flow harmonics of order $n$, $v_n$ scales with $v_2^{n/2}$,
consistent with quark coalescence. However, the ratio $v_4/v_2^2$ is
unexpectedly large. Some hadronic transport models are a factor of
2--3 lower than the data, but others achieve reasonable
agreement. However, hydrodynamic model calculations provide the best
predictions for $v_2$ compared with data. The characteristic
collectivity feature -- hadron mass dependence in the low $p_t$ region
-- is observed. Hydrodynamic models seem to work for minimum bias data
but not for centrality selected pion and anti-proton data. The
discrepancy for the central collision data may be due to nonflow
effects and fluctuations in the data, and for the peripheral
collisions from a failure of hydrodynamics. Perhaps, more work is
needed to improve the hydrodynamic fits, especially for the different
centrality bins, in order to make the case for early thermalization of
collisions at RHIC. Awaiting further theoretical input or explanation
are a number of STAR results, such as the large $v_2$ at high
$p_t$~\cite{STARhighPtV2Corr} and the $v_4$ observations. $v_4$ is
highly sensitive to initial conditions and the equation of state used
in hydrodynamic calculations, and therefore a challenge to all model
descriptions. The data were systematized with fits to a Blast Wave
model. The Blast Wave framework is capable of describing the large
volume of experimental data up to $p_t$ of 1 or 2 \GeVc\ using a
relatively small set of fit parameters in each centrality interval,
and the fit parameters are found to vary smoothly with centrality.

\section{\label{sec:acknow}Acknowledgments}       
We thank the RHIC Operations Group and RCF at BNL, and
NERSC at LBNL for their support. This work was supported
in part by the HENP Divisions of the Office of Science of the U.S.
DOE; the U.S. NSF; the BMBF of Germany; IN2P3, RA, RPL, and
EMN of France; EPSRC of the United Kingdom; FAPESP of Brazil;
the Russian Ministry of Science and Technology; the Ministry of
Education and the NNSFC of China; Grant Agency of the Czech Republic,
FOM and UU of the Netherlands, DAE, DST, and CSIR of the Government
of India; Swiss NSF; and the Polish State Committee for Scientific 
Research.

\appendix


\section{\label{sec:BWapp}Blast Wave equations} 
In both the surface shell and the filled elliptic cylinder cases the
transverse rapidity parameterization is extended to account for a
possible fourth harmonic anisotropy:
\begin{equation} \rho(\phi_b) = \rho_0 + \rho_2
\cos(2 \phi_b) + \rho_4 \cos(4 \phi_b), \label{eq:rho}
\end{equation}
where the flow magnitude and anisotropy are accounted for by the
$\rho_n$ parameters and $\phi_b$ is the azimuthal angle of the boost
source element defined with respect to the reaction plane, as
shown in Fig.~\ref{fig:ellipse}.

The distribution of source elements relative to $\phi_b$ in the case
of a surface shell is written including 2$^{\mathrm{nd}}$ and
4$^{\mathrm{th}}$ harmonic azimuthal anisotropy quantified by the
$s_2$ and $s_4$ parameters respectively:
\begin{equation}   
\Omega(\phi_b) = 1 + 2 s_2 \cos(2\phi_b) + 2 s_4 \cos(4\phi_b)
\end{equation}
When $s_2$ is positive more particles are boosted in-plane than
out-of-plane. In the case of a filled ellipse, the boost direction
($\phi_b$) is assumed to be perpendicular to the freeze-out surface,
which leads to a relationship between the space and boost azimuthal
angles of the emitted particles
\begin{equation}   
\tan(\phi_s) = (R_y/R_x)^2 \tan(\phi_b),
\label{eq:phi_s}
\end{equation}
with $R_x$ and $R_y$ the in-plane and out-of-plane radii,
respectively. For the $v_n$ analysis $R = (R_x^2 + R_y^2)^{1/2}$ is an
arbitrary radius, but when interferometry data are also fit, the units
become significant. The system is bounded within an ellipse such as
$\Omega(r,\phi_s) = \theta(\tilde{r}(\phi_s))$ with $\theta$ the step
function and
\begin{equation}   
\tilde{r}(\phi_s) = \sqrt{(r\cos(\phi_s)/R_x)^2+(r\sin(\phi_s)/R_y)^2}.
\end{equation}
In the filled ellipse case there is no explicit 2$^{\mathrm{nd}}$ and
4$^{\mathrm{th}}$ harmonic parameterization of the spatial
distribution of the particle emitting source because it is done
implicitly by the ellipse parameterization. A profile, linear in
transverse rapidity is used in the filled ellipse case:
\begin{equation}
\rho(r, \phi_b) = (\rho_0 + \rho_2 \cos(2 \phi_b) + \rho_4 \cos(4
\phi_b))~\tilde{r}(\phi_b).
\end{equation}

The flow Fourier coefficients are defined by
\begin{equation}
v_n = \la\cos[n(\phi_p - \Psi)]\ra,
\label{eq:vn}
\end{equation}
where $\phi_p$ is the azimuthal angle of the particle momentum.
Assuming a Boltzmann plus flow distribution and longitudinal boost
invariance, leads to the following expression for $v_n$:
\begin{widetext}
\begin{equation}   
v_n(p_t)= \frac
{\int_{-\pi}^{\pi} d\phi_s \int_0^{\infty} r dr \int_0^{\pi} 
d\phi_pK_1(\beta(r,\phi_b)) \cos(n\phi_p)
e^{\alpha(r,\phi_b)\cos(\phi_b-\phi_p)}
\Omega(r,\phi_s)
}
{\int_{-\pi}^{\pi} d\phi_s \int_0^{\infty} r dr \int_0^{\pi} 
d\phi_p K_1(\beta(r,\phi_b)) e^{\alpha(r,\phi_b)\cos(\phi_b-\phi_p)}
\Omega(r,\phi_s)
},
\label{solidblastwave}\end{equation}
with $\alpha(r,\phi_b)=(p_t/T)\sinh(\rho(r,\phi_b))$ and
$\beta_t(r,\phi_b)=(m_t/T)\cosh(\rho(r,\phi_b))$. The relation between
$\phi_b$ and $\phi_s$ is given by Eq.~(\ref{eq:phi_s}). All the
integrals are done numerically in the filled ellipse calculation in
order to preserve the possibility of computing interferometry
radii, even though the formula can be simplified to:
\begin{equation}   
v_n(p_t)= \frac
{\int_{-\pi}^{\pi} d\phi_b \int_0^{\infty} r dr 
K_1(\beta(r,\phi_b)) \cos(n\phi_b) I_n(\alpha(r,\phi_b)) \Omega(r,\phi_b)
}
{\int_{-\pi}^{\pi} d\phi_b \int_0^{\infty} r dr  
K_1(\beta(r,\phi_b)) I_0(\alpha(r,\phi_b)) \Omega(r,\phi_b)
}.
\label{shellblastwave}\end{equation} 
\end{widetext}
For the surface shell case the integral over $r$ is trivial. 


\section{\label{sec:derivV1}Derivation of the mixed harmonic event
plane method \vO} Following the discussion in \cite{STARv1v4}, we try
to reduce the nonflow contribution of the first harmonic signal,
$v_1$, by subtracting the contributions to the flow vector
perpendicular to the reaction plane from the component within the
reaction plane. As an estimate of the reaction plane we use the second
order event plane $\Psi_2$. Correlating the azimuthal angle of a
particle, $\phi$, with the first order event plane, $\Psi_1$, one then
obtains

\begin{widetext}
\begin{eqnarray}
\label{eq:v1_1}\langle \cos(\phi - \Psi_2) \cdot \cos(\Psi_1 - \Psi_2) - \sin(\phi -
\Psi_2) \cdot \sin(\Psi_1 -\Psi_2) \rangle &=& 
\langle \cos (\phi + \Psi_1 - 2\Psi_2 )\rangle = \\
\langle \cos(\phi - \Psi_{\mathrm{RP}}) \cdot \cos(\Psi_1 -
\Psi_{\mathrm{RP}}) \cdot\cos [2(\Psi_2 - \Psi_{\mathrm{RP}})] \rangle
&=&  \nonumber \\
\langle \cos(\phi - \Psi_{\mathrm{RP}})\rangle \cdot \langle
\cos(\Psi_1 - \Psi_{\mathrm{RP}}) \rangle \cdot \langle \cos [2(\Psi_2
- \Psi_{\mathrm{RP}})] \rangle &\equiv&
\label{eq:v1}v_1 \cdot \mathrm{Res}(\Psi_1) \cdot \mathrm{Res}(\Psi_2).
\end{eqnarray}
\end{widetext}
The factorization in Eq.~(\ref{eq:v1}) left-hand side is valid due to
the statistical independence of the three factors.  While the
resolution of the second order event plane, $\mathrm{Res}(\Psi_2)$,
can be obtained by calculating the square-root of the correlation of
two subevent planes, the resolution of the first order event plane,
$\mathrm{Res}(\Psi_1)$, can be calculated by considering
\begin{eqnarray}
\langle\cos [2(\Psi_1-\Psi_2)]\rangle &=& \nonumber \\
\langle\cos^2(\Psi_1-\Psi_{\mathrm{RP}})
\cdot \cos [2(\Psi_2-\Psi_{\mathrm{RP}})]\rangle &=& \nonumber \\ 
\langle\cos(\Psi_1-\Psi_{\mathrm{RP}})\rangle^2
\cdot \langle \cos [2(\Psi_2-\Psi_{\mathrm{RP}})]\rangle &=& \nonumber \\ 
\label{eq:res} \mathrm{Res}^2(\Psi_1)\cdot\mathrm{Res}(\Psi_2).
\end{eqnarray}
Combining Eqs.~(\ref{eq:v1}) right-hand side and (\ref{eq:res}) yields
\begin{displaymath}
\vO = \frac{\langle \cos (\phi + \Psi_1 - 2\Psi_2 )
  \rangle}{\sqrt{\langle\cos [2(\Psi_1-\Psi_2)]\rangle \cdot 
  \mathrm{Res}(\Psi_2)}} .
\end{displaymath}
This approach is similar to the three-particle correlation method
of Borghini, Dinh, and Ollitrault \cite{v1{3}}. One obtains their
result by replacing the event plane angles $\Psi_1$ and
$\Psi_2$ in the right-hand side of Eq.~(\ref{eq:v1_1}) by
emission angles of two particles~\cite{STARv1v4}.

Experimentally one wants to optimize the resolution of the second
order event plane by measuring it in a region $c$ where the signal of
$v_2$ is strong. This will be around mid-rapidity, preferentially. On
the other hand, the influence of nonflow can be reduced even further
by measuring the azimuthal angle of the particle in one subevent,
$\phi^a$, and correlating it to the first order event plane in the
other subevent, $\Psi_1^b$. These subevents might by chosen randomly,
or by dividing the acceptance into different regions in
pseudorapidity. Since only half of all particles are used to
determine each $\Psi_1^a$ and $\Psi_1^b$, the statistical
errors are increased by a factor of $\sqrt{2}$ compared to the
three-particle cumulant method $v_1\{3\}$.  The final
observable looks like this:
\begin{displaymath}
\vO = \frac{\langle \cos (\phi^a + \Psi_1^b - 2\Psi_2^c )
  \rangle}{\sqrt{\langle\cos(\Psi_1^a+\Psi_1^b-2\Psi_2^c)\rangle \cdot 
	\mathrm{Res}(\Psi_2^c)}} .
\end{displaymath}

In our case each particle azimuth was correlated to the first order
event plane determined in the other subevent within the FTPCs, and to
the second order event plane measured in the TPC.


\def\etal{\mbox{$\mathrm{\it et\ al.}$}}

\end{document}